\documentclass[a4paper,fleqn,usenatbib]{mnras}

\usepackage{newtxtext,newtxmath}

\usepackage[T1]{fontenc}
\usepackage{ae,aecompl}
\usepackage[dvipsnames]{xcolor}
\usepackage{ulem}

\usepackage{graphicx}	
\usepackage{amsmath}	

\newcommand{\kms}{{\rm {km\, s^{-1}}}}
\newcommand{\msun}{{\rm M_{\odot}}}

\definecolor{olive}{rgb}{0.5, 0.5, 0.0}
\definecolor{light-violet}{rgb}{0.93, 0.51, 0.93}

\title[IMBHs in merging stellar clusters]{Formation of super-massive black holes in galactic nuclei {\sc i}: delivering seed intermediate-mass black holes in massive stellar clusters
}
\author[A. Askar et al.]{
Abbas Askar,$^{1}$\thanks{E-mail: askar@astro.lu.se (AA)}
Melvyn B. Davies$^{1}$
and Ross P. Church$^{1}$
\\
$^{1}$Lund Observatory, Department of Astronomy, and Theoretical Physics, Lund University, Box 43, SE-221 00 Lund, Sweden
}

\date{Accepted 2021 January 12. Received 2021 January 12; in original form 2020 June 08}

\pubyear{2020}

\begin{document}
\label{firstpage}
\pagerange{\pageref{firstpage}--\pageref{lastpage}}
\maketitle

\begin{abstract}
Supermassive black holes (SMBHs) are found in most galactic nuclei. A significant fraction of these nuclei also contain a nuclear stellar cluster (NSC) surrounding the SMBH. In this paper, we consider the idea that the NSC forms first, from the merger of several stellar clusters that may contain intermediate-mass black holes (IMBHs). These IMBHs can subsequently grow in the NSC and form an SMBH. We carry out $N$-body simulations of the simultaneous merger of three stellar clusters to form an NSC, and investigate the outcome of simulated runs containing zero, one, two and three IMBHs.
We find that IMBHs can efficiently sink to the centre of the merged cluster. If multiple merging clusters contain an IMBH, we find that an IMBH binary is likely to form and subsequently merge by gravitational wave emission. We show that these mergers are catalyzed by dynamical interactions with surrounding stars, which systematically harden the binary and increase its orbital eccentricity. The seed SMBH will be ejected from the NSC by the recoil kick produced when two IMBHs merge, if their mass ratio $q\gtrsim 0.15$.
If the seed is ejected then no SMBH will form in the NSC. This is a natural pathway to explain those galactic nuclei that contain an NSC but apparently lack an SMBH, such as M33. However, if an IMBH is retained then it can seed the growth of an SMBH through gas accretion and tidal disruption of stars.
\end{abstract}


\section{Introduction}\label{sec:intro}

The nuclei of most galaxies with stellar masses larger than $10^{8} \, \msun$ contain either a nuclear star cluster (NSC), a supermassive black hole (SMBH), or both. NSCs are massive and dense stellar clusters that reside in the nuclei of their host galaxies. They typically have effective radii of a few parsecs with central densities of up to $\rm 10^{7} \, \msun \, pc^{-3}$ \citep{Georgiev2016,schodel2018}. About 80 per cent of galaxies of all morphological types that have stellar masses in the range $10^8$ to $10^{10} \, \msun$ harbour an NSC \citep[see][and references therein]{Neumayer2020}. In many galaxies, including the Milky Way and M31, an NSC coexists with an SMBH \citep{seth2008,Neumayer2012,Neumayer2020}. Some galaxies however, such as M33, contain an NSC \citep{kormendy1993,gordon1999} but show no evidence for the presence of an SMBH in their nuclei \citep{Merritt2001,Gebhardt2001}. Typically, elliptical galaxies with stellar masses greater than $10^{11} \, \msun$ have low nucleation fractions and contain only an SMBH at their centre \citep{cote2006,graham2009,Seth2019}. Both NSCs and SMBHs follow tight correlations with the properties of their host galaxies \citep{Ferrarese2006,leigh2015,capuzzo2017} and this suggests that the formation and subsequent growth of both NSCs and SMBHs could be closely related \citep{antonini2015,Neumayer2020}. 

Two main scenarios for the formation of NSCs have been proposed. 
NSCs may have formed from the merger of infalling stellar clusters that aggregate to the galactic centre due to dynamical friction \citep{tremaine1975,cd1993,lotz2001,oh2000,agarwal2011,tsatsi2017} or they may have formed through in-situ star formation triggered by high gas densities in the galactic nuclei \citep{loose1982,mihos1994,milosavljevi2004,nayakshin2007,aharon2015}. 
Observations of NSCs show that they contain stars with diverse ages and metallicities and it is likely that both cluster infall and in-situ star formation contribute to the growth of NSCs \citep{antonini2015,guillard2016,Neumayer2020,2020arXiv200902335D,2020arXiv200902328A}.

SMBHs have masses in the range $10^{6} - 10^{10} \, \msun$ and their formation and growth remains an open question \citep{latif2016}. It is likely that SMBHs grew from smaller seed black holes (BHs), however, the exact seeding mechanism and the masses of these seed BHs are not known. These seed BHs could have originated from the evolution of massive stars or they may have formed with large initial masses of the order $10^{5} \, \msun$ \citep{volonteri2010,johnson2013,greene2019rev}. In the former case, BHs with masses in the intermediate-mass range of $10^{2}-10^{5} \, \msun$ ought to have existed in order to seed SMBHs. It has been postulated that such intermediate-mass BHs (IMBHs) could form through dynamical processes in dense stellar clusters such as globular clusters or young massive clusters (see Section \ref{subsec:imbh-formation}). Due to uncertainties in their formation mechanism, it is not clear which forms first, the NSC or an SMBH. If an SMBH grows from lower-mass seed BHs in dense environments then it is possible that the formation of an NSC precedes the formation of an SMBH \citep{Devecchi2009,Davies2011,Miller2012,Neumayer2012,gnedin2014}.

Given that an NSC may form due to the inspiral and merger of smaller stellar clusters, it is likely that some of the merging clusters may already have formed a seed IMBH, which could then be delivered to the NSC \citep{ebisuzaki2001,kim2004,spz2006,devecchi2012,petts2017,davies2020}. Once an IMBH is delivered to the NSC, it can become a seed SMBH and continue to grow through tidal captures/disruption of stars \citep{stone2017,alexander2017,boekholt2018,sedda2019b}
or through continued gas accretion which can also contribute to the growth of the NSC \citep{Devecchi2009,Davies2011, guillard2016,paucci2017,natarajan2020,das2020}. In the latter case, a seed IMBH of $10^{4}$ $\msun$ can grow up to $10^{9}$ $\msun$, if it is able to accrete at the Eddington rate for 500 Myr \citep{madau2014}. BHs can grow rapidly if they have low spin values following episodes of chaotic accretion \citep{king2006,king2008}.
It is also likely that multiple IMBHs are delivered to the growing NSC from more merging clusters \citep{ebisuzaki2001,AMB2014}.
This may result in the formation of a binary IMBH which could merge by emitting gravitational waves \citep{amaro2006,Tamfal2018,rasskazov2019,SeddaAMB2019,wirth2020}. Provided that the gravitational wave (GW) recoil kick from the merger of two IMBHs is less than the escape speed of the NSC, the merged IMBH can be retained in the NSC and can grow into an SMBH \citep{amaro2006,gurkan2006,as2007,lisa2017,SeddaAMB2019}. However, if the gravitational recoil kick is larger than the escape speed of the cluster then the merged IMBH will be ejected from the NSC and the seed SMBH will be lost. Dynamical ejection of one or more IMBHs can also occur through binary-single encounters involving three IMBHs. Therefore, these interactions could also potentially remove a seed SMBH from an NSC.

Several studies have numerically investigated the formation of an NSC by merging stellar clusters. \citet{hartmann2011} carried out collisionless \textit{N}-body simulations of stellar clusters merging at the centre of a galactic disc to form an NSC that had properties comparable to observed NSCs. \citet{cd2006,cd2008} carried out simulations of clusters infalling in a galactic centre and showed that they can survive strong tidal interactions and merge to form dense NSCs. \citet{Antonini2012} carried out \textit{N}-body simulations to follow the successive inspiral and merger of twelve globular clusters in an initial setup which included a low-density nuclear stellar disc and an SMBH similar in mass to Sagittarius A*. They found that the NSC produced from these merging clusters had properties similar to those of the Milky Way NSC, and concluded that nearly half of the stars in the Milky Way NSC originated from merging clusters while the rest probably formed in-situ. \citet{AMB2014} used the same initial set-up as \citet{Antonini2012} but included IMBHs at the centre of their merging clusters. They found that the IMBHs inspiral to the core of the NSC where they interact with surrounding stars and each other. The inclusion of IMBHs shortens the relaxation time of the NSC and also increases the rate of tidal disruption events.
\citet{ArcaSedda2015} also carried out simulations of merging stellar clusters with the goal to model the future evolution of the galactic nucleus in Henize 2--10. With the exception of one run, they also had an existing SMBH in their galactic nuclei. They found that an NSC forms in all their models and in the simulated run without an SMBH, the most massive infalling clusters lose less mass as they decay towards the galactic centre. \citet{sedda2017} used \textit{N}-body simulations to investigate the role of galaxy structure in determining the density distribution of an NSC and its implications for SMBH seeding.

Semi-analytical models have also been developed to investigate  the formation of NSCs through mergers of globular clusters that undergo dynamical friction \citep{antonini2013,gnedin2014,sedda2014}. The results from these models are able to reproduce some of the observed properties of NSCs \citep{Neumayer2020}. Using models for galaxy formation that included prescriptions for NSC formation and dynamical heating from massive BHs, \citet{antonini2015} were able to account for low nucleation fractions in massive elliptical galaxies \citep{Neumayer2012,sedda2016b}.

In this work, we focus on the formation of an NSC through mergers of stellar clusters. We carry out several \textit{N}-body simulations to follow the final stages of an idealized merger of three stellar clusters with no existing SMBH. Firstly, we merge clusters that only contain stars (Section \ref{section2:merging clusters}); then we include various combinations of IMBHs at the centre of the merging clusters to investigate how IMBHs would evolve within the merged cluster. In Section \ref{section3-one-imbh}, we describe runs in which there is a single IMBH in the three merging clusters and also discuss various mechanisms by which IMBHs can form in dense star clusters (Section \ref{subsec:imbh-formation}). 
In Sections \ref{2i-runs} and \ref{sec4:3i-runs}, we present results from runs in which the initial set-up contains two and three IMBHs. We find that in these runs, we can form a binary IMBH that hardens and becomes more eccentric as it interacts with surrounding stars \citep{quinlan1990,sesana2006,baumgardt2006,iwasawa2011}. By reaching high orbital eccentricities, these binaries can efficiently merge due to GW emission within several hundred Myr. The evolution of the orbital parameters of these binaries are discussed in Appendix \ref{ross-section}.

In Section \ref{imbh-removal}, we describe important processes such as GW recoil kicks (Section \ref{sec:4-gw-recoil-kicks}) and strong scattering encounters (Section \ref{sec4:binsin-interaction}) between IMBHs that can influence the retention of seed SMBHs in NSCs. We argue that there are two routes that endow galaxies with an NSC but no SMBH. Firstly, they may never have had a seed IMBH; i.e. none of the clusters that combine to form the NSC contained an IMBH.  The second possibility is that the seed SMBH may have been ejected whilst it was still growing. If an IMBH is retained or already exits in an NSC, then it can grow to become an SMBH. In a subsequent paper, we use the results presented here to model the retention and growth of seed SMBHs in a population of galaxies where the NSC is constructed through the merger of stellar clusters.

\section{Simulating Merging Stellar Clusters}\label{section2:merging clusters}

We simulate the final stages of an idealized merger of three stellar clusters to form an NSC. In Section \ref{sec2:model-description}, we describe the details of our simulation set-up for three merging clusters and briefly describe how these clusters evolve in Section \ref{sect:noIMBH}. In the runs presented in this paper, there is no existing SMBH and there is no galactic tidal field.

\subsection{\textit{N}-body simulations: Set-up and Initial Conditions}\label{sec2:model-description}

We use a series of direct \textit{N}-body simulations of the merger of three stellar clusters. We consider a range of cases where various combinations of the three merging clusters initially contain an IMBH at their centre. The simulations were carried out using the \textsc{NBODY6++GPU} \citep{wang2015} code. This is an extension of the well-known \textsc{NBODY6} code developed by \citet{aarseth1999,aarseth2003gravitational} and its GPU accelerated version \textsc{NBODY6GPU} code \citep{nitadori2012} for simulating star cluster evolution. 

\textsc{NBODY6++GPU} uses the fourth-order Hermite integration method together with hierarchical block time steps. In order to speed up force calculations, the code uses \citet{ahmadcohen1973} neighbor scheme that separates integration of  regular and irregular forces using large time-steps and small time-steps respectively.  \textsc{NBODY6++GPU} combines GPU acceleration with MPI parallelization so that the code maybe used across multiple nodes on computer clusters. The code can accurately treat binary systems and close encounters through the \citet{ks1965} and chain regularization \citep{mikkola1993} algorithms. It also contains prescriptions for stellar and binary evolution based on the \textsc{SSE} and \textsc{BSE} codes \citep{hurley2000,hurley2002}. 

Using \textsc{NBODY6++GPU} we generated three initial models for stellar clusters comprising 50\,000,  30\,000 and 15\,000 particles. Each of these was a Plummer model consisting of single stars with masses between 0.5 and 2 $\rm M_{\odot}$. Within this range, the  zero-age-main sequence (ZAMS) masses were sampled using a power law initial mass function with $\alpha = 2.3$ \citep{kroupa2001}. Initial ZAMS radii for the stars were generated for a metallicty of $Z=0.001$ ([Fe/H] = $-1.301$). The initial half-mass radii of the clusters were 2.4 pc with a Plummer scale length of 1.82 pc. Table \ref{tab:three-ini-clustrs} contains the properties of the three modelled clusters. For each of the three stellar clusters, we also generated models that, in addition to stars, contained a central IMBH of 1000, 500, 200, or 100 $\rm M_{\odot}$.

\begin{table*}
\centering
  \caption{In this table, the initial properties of the three individual merging clusters are provided} 
  \label{tab:three-ini-clustrs}
\begin{tabular}{|c|c|c|c|c|c|}
\hline
\textbf{Model} & {\textbf{Number}} & {\textbf{Initial Mass}} & \textbf{Half-mass} & \textbf{Plummer Scale} & \textbf{Half-Mass} \\ & & \textbf{[$\rm M_{\odot}$}] & \textbf{Radius [pc]} & \textbf{Radius [pc]} & \textbf{Relaxation Time [Myr]}  \\ \hline
\textbf{Central} & $\rm 5.0 \times 10^{4}$ & $\rm 4.42 \times 10^{4}$ & 2.40 & 1.84 & 160 \\
\textbf{First Infalling} & $\rm 3.0 \times 10^{4}$ & $\rm 2.65 \times 10^{4}$ & 2.40 & 1.84 & 130 \\
\textbf{Second Infalling} & $\rm 1.5 \times 10^{4}$ & $\rm 1.33 \times 10^{4}$ & 2.40 & 1.84 & 99\\
\hline
\end{tabular}
 \end{table*}

\begin{figure}
	\includegraphics[width=\columnwidth]{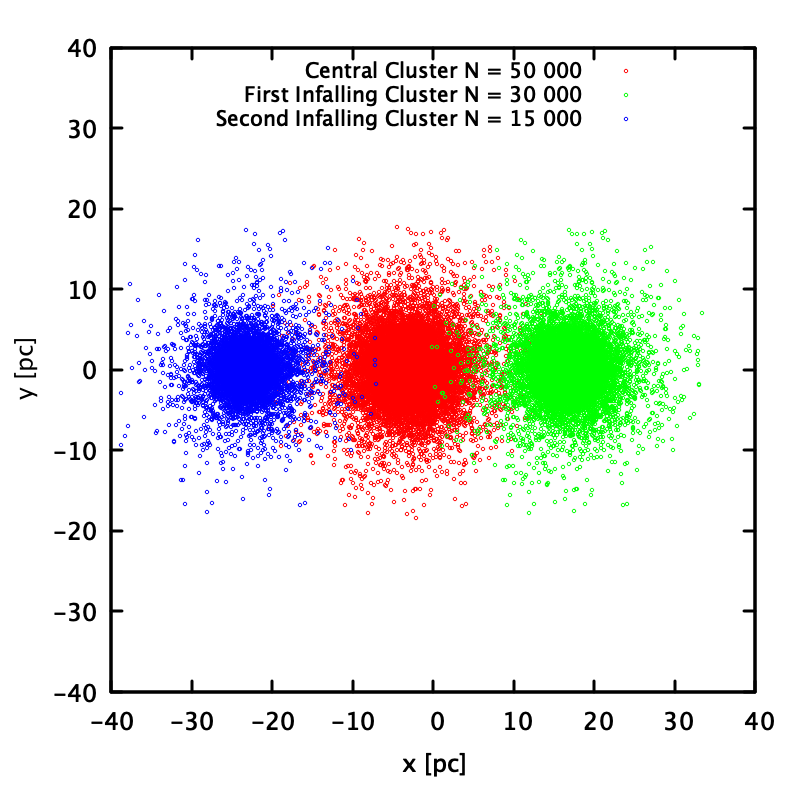}
	\caption{$X-Y$ particle plot showing the initial set-up for the three merging clusters described in Section \ref{sec2:model-description}. The origin is centred on the global centre of mass. Red particles are stars that belong to the cluster placed in the middle with 50\,000 stars, the green particles belong to the cluster with 30\,000 stars which is offset in the $X$ direction from the centre of the central cluster by 20 pc and the blue particles belong to the cluster with 15\,000 stars which is offset in the $X$ direction by -20 pc.}
    \label{fig:t0-particle-plot}
\end{figure}

Using these three initial models, we set-up an initial model which is meant to represent the final stage of the merger of three stellar clusters. We place the cluster with 50\,000 particles in the centre, the cluster with 30\,000 particles is offset in the $X$ direction by 20 pc from the centre, and the cluster with 15\,000 particles is offset in the $X$ direction by -20 pc. In order to place the clusters with 30\,000 particles and 15\,000 particles on sub-circular orbits around the central cluster, we added offsets to the $Y$ component of the initial velocities of -2.8 and 1.4 $\kms$ respectively. Fig.  \ref{fig:t0-particle-plot} is a particle plot showing the initial $X-Y$ positions of the three clusters in a coordinate system centred on the global centre of mass. The two 30\,000 and 15\,000 particle clusters were placed sufficiently close to the central cluster so that they both fill their Roche-lobe and merge with the central cluster on a dynamical timescale. While the number of stars and the total stellar mass in these merging clusters is smaller than typical NSCs, the approach taken in this work allows us to capture IMBH dynamics and the effect of close encounters involving stellar mass stars on IMBH binary evolution. This is particularly important for the results of the runs presented in Section \ref{2i-runs} and \ref{sec4:3i-runs}.


\vspace{0.3cm}

In this section, we present a run where the merging clusters contain only stars. We then add IMBHs to this initial set-up in order to differentially measure the effects of IMBH inclusion. In Section \ref{section3-one-imbh}, we describe the runs with one IMBH. In Sections \ref{2i-runs} and \ref{sec4:3i-runs}, we describe runs that contain two and three IMBHs. The runs have been named in the format X.i, where X indicates the number of IMBHs in that run and i is the run number. So 0.1, is the run which contains no IMBH, 1.2 is the second run with initially one IMBH. In total we have eleven simulations comprising one run with no IMBH (0.1), three runs with one IMBH (1.i), four runs with two IMBHs (2.i), and three runs with three IMBHs (3.i). 

\begin{figure*}
	 \includegraphics[width=0.48\linewidth]{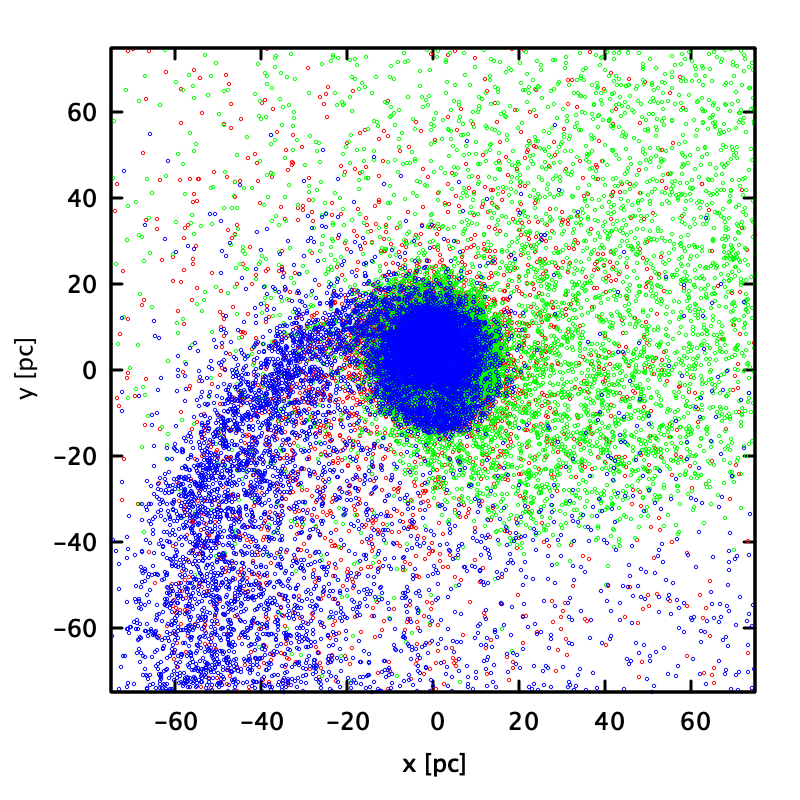}
	  \includegraphics[width=0.48\linewidth]{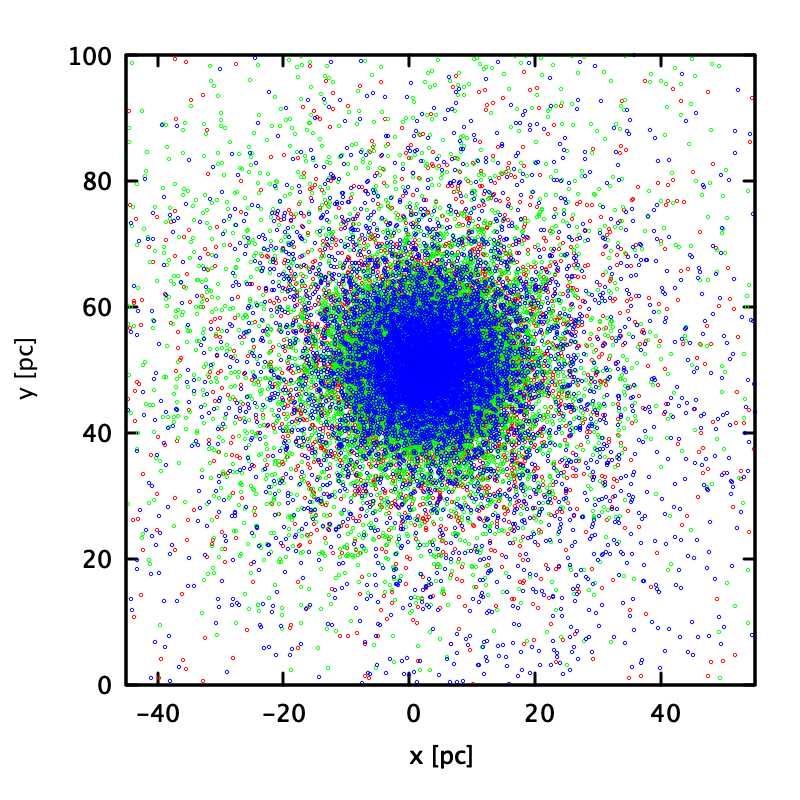}
	\caption{Particle plots in the $X-Y$ plane for model 0.1, which contains no IMBHs. The different colours represent stars from the three initial clusters as shown in Fig.~\ref{fig:t0-particle-plot}. The left panel shows the clusters spiralling together at 28.8 Myr. The right panel shows the resulting single, merged cluster at 360 Myr. The core of the cluster appears blue owing to a plotting artefact: the core is a mixture of stars from all the three stellar clusters.}
    \label{fig:0.1-particle}
\end{figure*}

\subsection{Evolution of merging clusters without IMBHs}
\label{sect:noIMBH}

Within the first few time steps, the two infalling clusters begin to merge into the central cluster. As the merger proceeds, the infalling clusters produce large tidal tails of stars, some of which gradually return and fall in to the central cluster. We simulated the model up to 375 Myr, by which time all the three clusters had merged into a single cluster with a dense core. In Fig.~\ref{fig:0.1-particle}, we show particle plots at 28.8 Myr and 360 Myr. At 28.8  Myr, the tidal tails of stars from the infalling star clusters with 30\,000 and  15\,000 stars can be seen in green and blue respectively. At 360 Myr, we find that the final merged cluster now has a core containing stars from all the three merged clusters and there are no visible tidal tails in the vicinity of the cluster.

\begin{figure}
	\includegraphics[width=\columnwidth]{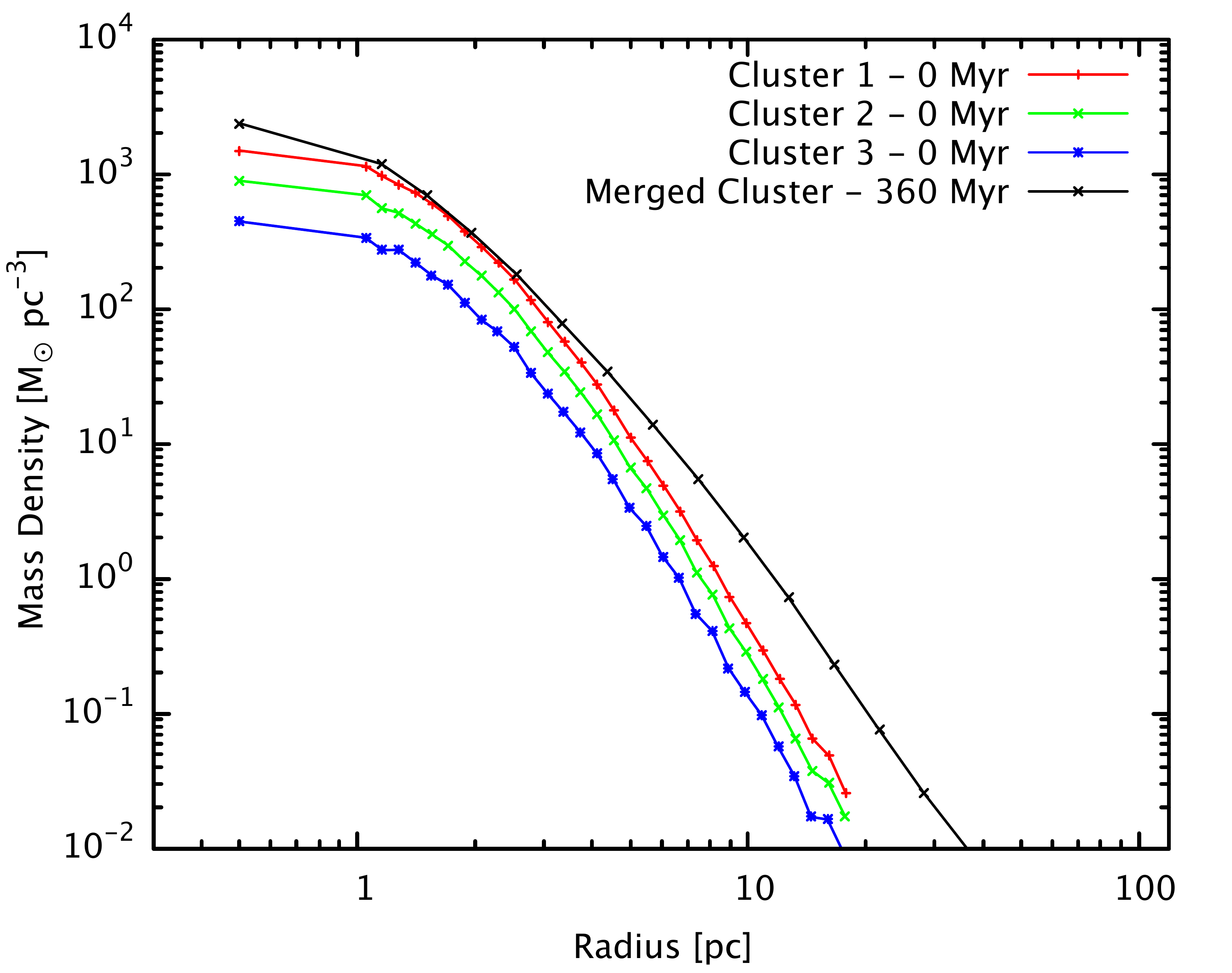}
	\caption{Density profiles for the three merging clusters at $t=0$ and the merged cluster at 360 Myr (black line). The red coloured line shows the initial density profile for the central cluster with 50\,000 stars, the green line corresponds to the more massive infalling cluster that contains 30\,000 stars, and the blue line corresponds to the cluster with 15\,000 stars. The merged cluster at 360 Myr is more radially extended and has a denser core than the three initial merging clusters.}
    \label{fig:0imbh-density}
\end{figure}

In Fig.~\ref{fig:0imbh-density}, we compare the density profile of the merged cluster at 360 Myr to the initial density profiles of the three individual merging clusters in Fig.~\ref{fig:0imbh-density}. The merged cluster at 360 Myr has a slightly larger central density than the initial clusters. This increase in density of the merged cluster due to gravitational encounters is also reported by \citet{Antonini2012} and \citet{AMB2019}.

\section{Merging star clusters with one IMBH}\label{section3-one-imbh}

\subsection{IMBH formation in star clusters}\label{subsec:imbh-formation}

It has been hypothesized that IMBHs of $10^{2} - 10^{4}\,\rm M_{\odot}$ may form in the dense environments of stellar clusters \citep[see][and references therein]{greene2019rev}, either through runaway mergers of massive stars that may lead to the formation of a seed IMBH \citep{SPZ02,freitag2006,Giersz15,mapelli2016,gieles2018,reinoso2018,tagawa2020}, or through the gradual growth of stellar-mass BHs via mergers with other BHs or stars \citep{Miller02,Giersz15,haster2016,rizzuto2020}. Seed BHs of $\rm 10^{2} - 10^{3}\,\rm M_{\odot}$  can grow to larger masses through tidal capture and disruption events in dense clusters \citep{stone2017,alexander2017,sakurai2019}. Moreover, stellar-mass seed BHs can also grow by accretion of gas in primordial, gas-rich, massive stellar clusters \citep{vesperini2010,leigh2013}. Observationally, identifying the presence of an IMBH in a stellar cluster is challenging \citep{noyola2008,mezcua2017,devita2017,zocchi2019,aros2020} and kinematic studies searching for the presence of IMBHs in Galactic globular clusters have yielded inconclusive results \citep{lutzgendorf13,lanzoni13,kiziltan2017,mann2019,baumgardt2019}. Searching for accretion signatures of an IMBH in Galactic globular clusters using radio observations has also not found any evidence for IMBHs \citep{strader2012,tremou18}. \citet{lin18} identified an IMBH candidate in an extragalactic massive star cluster through an X-ray flare that originated from a tidal disruption event \citep{lin2020}. Additionally, a few low mass SMBHs have also been observed in galactic nuclei. \citet{baldassare2015}, used optical and X-ray observations to identify a BH with a mass of $5 \times 10^{4} \, \msun$ in the dwarf galaxy RGG 118. \citet{Nguyen2019} constrained the mass of a BH at the centre of the dwarf elliptical galaxy NGC 205 to be less than $7 \times 10^{4} \, \msun$. Recent observations of molecular gas streams in the Galactic centre have also identified possible IMBH candidates \citep{takekawa2019,takekawa2020}. The detection of gravitational waves from the merger of an $\sim 85 \ \msun$ and a $\sim 66 \ \msun$ BH (GW190521) by the LIGO/Virgo collaboration \citep{abbott2020} resulted in the formation of a $\sim 142 \ \msun$ BH. This is the first clear detection of a low-mass IMBH and it has been suggested that such massive merging binary BHs can form in star clusters \citep{dicarlo2020,samsing2020,fragione2020b,martinez2020,kremer2020,sedda2020-gw,shaw2020}.

Monte Carlo simulations of globular cluster models by \citet{Giersz15} using the \textsc{mocca} code \citep{hypki2013,giersz2013} showed that an IMBH may grow in a globular cluster through direct collisions and dynamical interactions between a BH and other objects. In \citet{Giersz15}, two scenarios for IMBH formation are described, the ``fast'' and ``slow'' scenarios. In  the fast scenario, an IMBH of $\rm 10^{2} - 10^{4}\,\rm M_{\odot}$ forms in initially dense globular cluster models -- those with central densities of $\rm \gtrsim 10^{7} \, M_{\odot} \, pc^{-3}$. In a significant fraction of these simulations, the IMBH formation occurs within tens of Myr due to runaway mergers between massive main-sequence stars \citep{quinlan1990,lee1993,SPZ02} leading to the formation of a massive star which absorbs a stellar-mass BH \citep{rizzuto2020}.
These models assume that all of the mass of the main-sequence star is transferred to the BH in such mergers, facilitating the formation of the seed IMBH. Additional simulations from \citet{Giersz15} in which only 25 per cent of the main sequence star mass is accreted by the BH in the case of such mergers also lead to the formation of an IMBH. In the ``slow'' scenario, a stellar-mass BH grows gradually through mergers with other BHs or through tidal disruption of stars during core collapse. In these models, the core collapse occurs after few to several Gyr of cluster evolution in models that are initially not too dense \citep{Sedda2019}. The fraction of these simulated models that form an IMBH with a mass of at least $100 \, \msun$ correlates with the initial mass of the cluster. About 20 per cent of clusters with an initial mass of around $10^5 \, \msun$ form an IMBH. This fraction increases to about 50 per cent for models with an initial mass of around $10^6 \, \msun$.

\begin{figure}
	\includegraphics[width=\columnwidth]{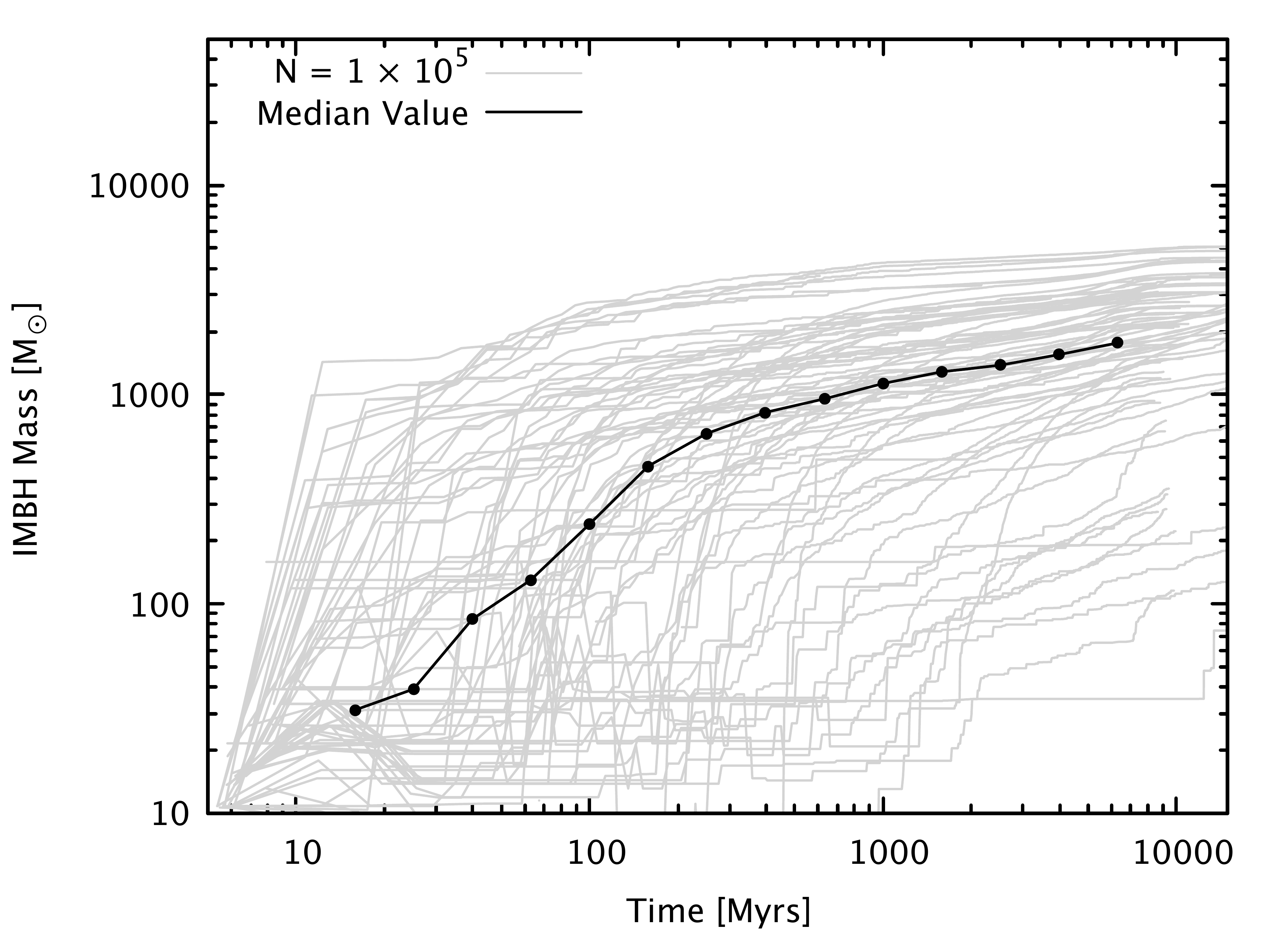}
	\caption{Masses of IMBHs as functions of time produced in clusters initially containing $1.0 \times 10^{5}$ stars.  The results are taken from the Monte-Carlo $N$-body code \textsc{mocca} and the \textsc{mocca}-Survey Database I project \citep{askar2017}. Each grey line represents a different simulation. The black line shows the median IMBH mass over the model set.}
    \label{fig:imbh-growth}
\end{figure}

In Fig.~\ref{fig:imbh-growth}, we show the evolution of the most massive BHs in \textsc{mocca} simulations of stellar cluster models that formed a BH larger than $100 \, \msun$. These models had initially $\rm 10^{5}$ stellar systems (binary systems + single stars) and different initial parameters. Details of the initial model set-up for the \textsc{mocca}-Survey Database I can be found in \citet{askar2017}. The median IMBH mass in these clusters after 100 Myr of evolution is a few hundred solar masses. The number of stars and the total mass in these simulations depend the initial binary fraction which varied between 5 and 95 per cent. The initial masses of the clusters plotted in \ref{fig:imbh-growth} are between $5.85 - 9.12 \times 10^4$ $\msun$. For these models, the IMF is sampled between 0.08 and $100\,\msun$.

\subsection{Merging clusters that contain 1 IMBH (1.i runs)}

We simulated three runs which had the same set-up as the run described in Section~\ref{sect:noIMBH}, but with an IMBH in the centre of one of the merging clusters. The IMBH mass and the cluster to which we added it are described in Table~\ref{tab:1-imbh-models}. In these runs, the IMBH mass was taken to be a reasonable fraction of the total cluster mass. We consider IMBHs of 1000 $\msun$, 500 $\msun$ and 200 $\msun$ that are consistent with IMBH masses after few hundred Myr in \textsc{mocca} simulations with initially 40\,000 to $\rm 10^{5}$ objects \citep{Sedda2019}.

\begin{table}
\centering
  \caption{In this table we describe the runs that contain one IMBH. The first column contains an identifier for the run, the second column contains the time up to which the simulation was run in Myr. The third column contains the total mass of the merging clusters. Columns 4 to 6 give the IMBH mass in the run and also describe the position of the IMBH in the initial set-up. $\rm IMBH_{1}$ indicates that the IMBH was initially at the centre of the cluster with 50\,000 stars, $\rm IMBH_{2}$ is initially at the centre of the cluster with 30\,000 stars, and $\rm IMBH_{3}$ is at the centre of the cluster with 15\,000 stars. Column 7 gives the time it takes the BH to sink to the central part of the merged cluster (within 0.25 pc from the centre of mass).}
  \label{tab:1-imbh-models}
\resizebox{\columnwidth}{!}{%
\begin{tabular}{|l|c|c|c|c|c|c|}
\hline
\textbf{Run} & {\textbf{Time}} & {\textbf{Total Mass}} & \textbf{$\rm IMBH_{1}$} & \textbf{$\rm IMBH_{2}$} & \textbf{$\rm IMBH_{3}$} & {\textbf{$\rm t_{sink}$}}  \\ & [Myr] & [$\rm M_{\odot}$] & [$\rm M_{\odot}$] & [$\rm M_{\odot}$] & [$\rm M_{\odot}$] & [Myr]  \\ \hline
1.1 & 250 & $\rm 8.50 \times 10^{4}$ & 1000 & 0 & 0 & 35 \\
1.2 & 357 & $\rm 8.45 \times 10^{4}$ & 0 & 500 & 0 & 61\\
1.3 & 551 & $\rm 8.42 \times 10^{4}$ & 0 & 0 & 200 & 48\\ \hline
\end{tabular}
}
 \end{table}

\begin{figure}
	\includegraphics[width=\columnwidth]{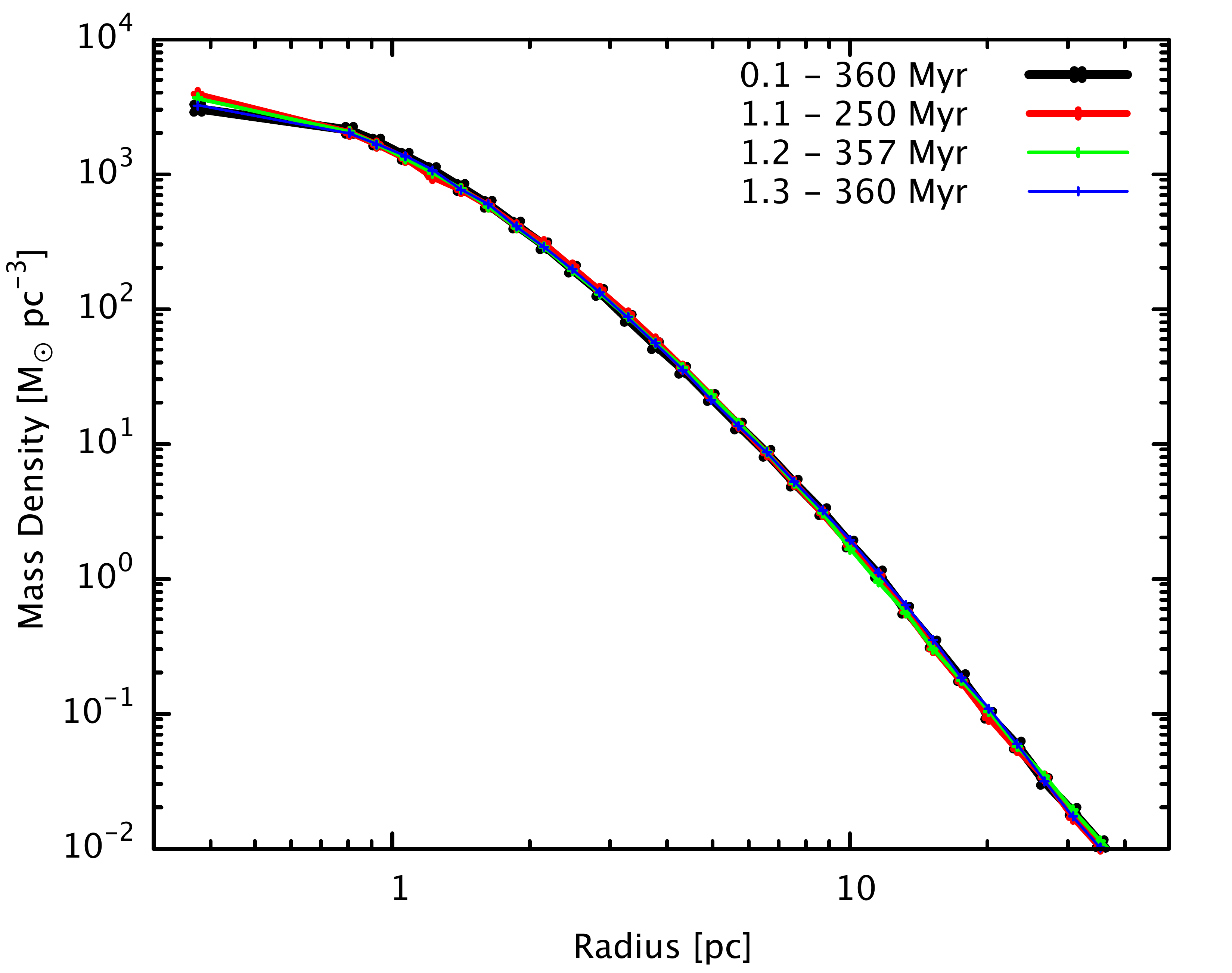}
	\caption{Mass density profiles for the merged clusters in runs 1.1 (red), 1.2 (green) and 1.3 (blue) and run 0.1 (black). The profiles were plotted at about 360 Myr from the start of the simulation for runs 1.2, 1.3 and 0.1. For cluster 1.1, the plot was made for the snapshot at 250 Myr (see Table \ref{tab:1-imbh-models}). All the four merged clusters have a similar density profile.}
    \label{fig:1imbh-density}
\end{figure}

We find that runs 1.1,1.2 and 1.3 evolve in a similar way to run 0.1. Since the mass of the IMBH in run 1.i is much smaller than the total mass of the merging clusters, presence of the IMBH
does not significantly affect the cluster evolution. Hence,
the properties and appearance of the merged cluster are the
same as run 0.1. To illustrate this, we show the mass density profile of runs 1.i and 0.1 in Fig. \ref{fig:1imbh-density}. For runs 0.1, 1.2 and 1.3, the profiles are plotted at about 360 Myr, while for run 1.1 the profile is shown for the last simulation snapshot at 250 Myr. The density profiles for the merged clusters in runs 1.i are similar and are also in agreement with run 0.1. The only differences are in the central density value which includes the IMBH mass.

In run 1.1, the IMBH was already at the centre of the most massive cluster that we placed in the middle while in runs 1.2 and 1.3 it was brought along with the two less massive infalling clusters. In all the runs, the IMBH ends up in the cluster centre due to mass segregation. It has been shown that the orbital decay time of the IMBH in an NSC depends on its mass and the density of surrounding stars \citep{sedda-gualandris2018}. Since $\rm M_{\star} \ll M_{\rm IMBH}$ in our runs, dynamical friction is efficient at bringing the IMBHs to centre of the merged cluster. In run 1.2, the $500 \, \msun$ IMBH, that was initially at the centre of the infalling cluster with 30\,000 stars, arrives at the centre of the merged cluster in about 20 Myr and settles within 0.25 pc from the centre of mass within 60 Myr. In run 1.3, a $200 \, \msun$ IMBH was placed at the centre of the infalling cluster with 15\,000 stars and it arrives at the centre of the merging cluster in about 38 Myr and settles within 0.25 pc of the centre of mass within 50 Myr. These runs show that the IMBHs brought along by merging stellar clusters can effectively sink to the centre of an NSC within a few tens of Myr from the beginning of the cluster merger.

One of the underlying assumptions in these runs is that the individual star clusters had evolved for about 100 Myr before they merged with each other. This assumption has implications on the location where the clusters may have been born in order for them to effectively sink to the inner 20 pc of a galaxy on a timescale of about a 100 Myr and have had sufficient time to form an IMBH. There is observational evidence for the presence of young clusters with masses larger than $10^{5} \ \msun$ in the innermost 150 pc of a few dwarf starburst galaxies and the inferred dynamical friction time for these clusters is $\sim$ 0.1-1 Gyr \citep{Nguyen2014,antonini2014,ArcaSedda2015}.

\section{Simulations containing two IMBHs (2.i runs)}
\label{2i-runs}

If NSCs are built up from multiple merging stellar clusters then it is possible that some of these clusters have formed IMBH in their centres through the mechanisms described in Section \ref{subsec:imbh-formation}. In that case, multiple IMBHs may be delivered to the galactic nucleus by merging stellar clusters. To check how multiple IMBHs evolve within the merged clusters, we also carried out runs in which we placed two or three IMBHs at the centre of the merging clusters. The evolution of the runs containing two IMBHs are described in this Section.

Using the same initial setup for three merging clusters described in Section~\ref{sec2:model-description},
we set-up four runs in which we placed IMBHs at the centre of two of the three merging clusters. In three of these runs (labelled 2.1 to 2.3), the central cluster with 50\,000 particles contains an IMBH of 1000 $\rm M_{\odot}$ and the first infalling cluster, with 30\,000 particles, contains IMBHs of masses 500 $\rm M_{\odot}$ (run 2.1), 100 $\rm M_{\odot}$ (run 2.2) and 200 $\rm M_{\odot}$ 
(run 2.3). In the run labelled 2.4, the central cluster does not contain an IMBH; instead, the infalling clusters with 30\,000 and 15\,000 particles contain IMBHs of 500 $\rm M_{\odot}$ and 200 $\rm M_{\odot}$ respectively. The set-up of these runs is summarised in Table~\ref{tab:2-imbh-models}.
 
\begin{figure*}
	 \includegraphics[width=0.48\linewidth]{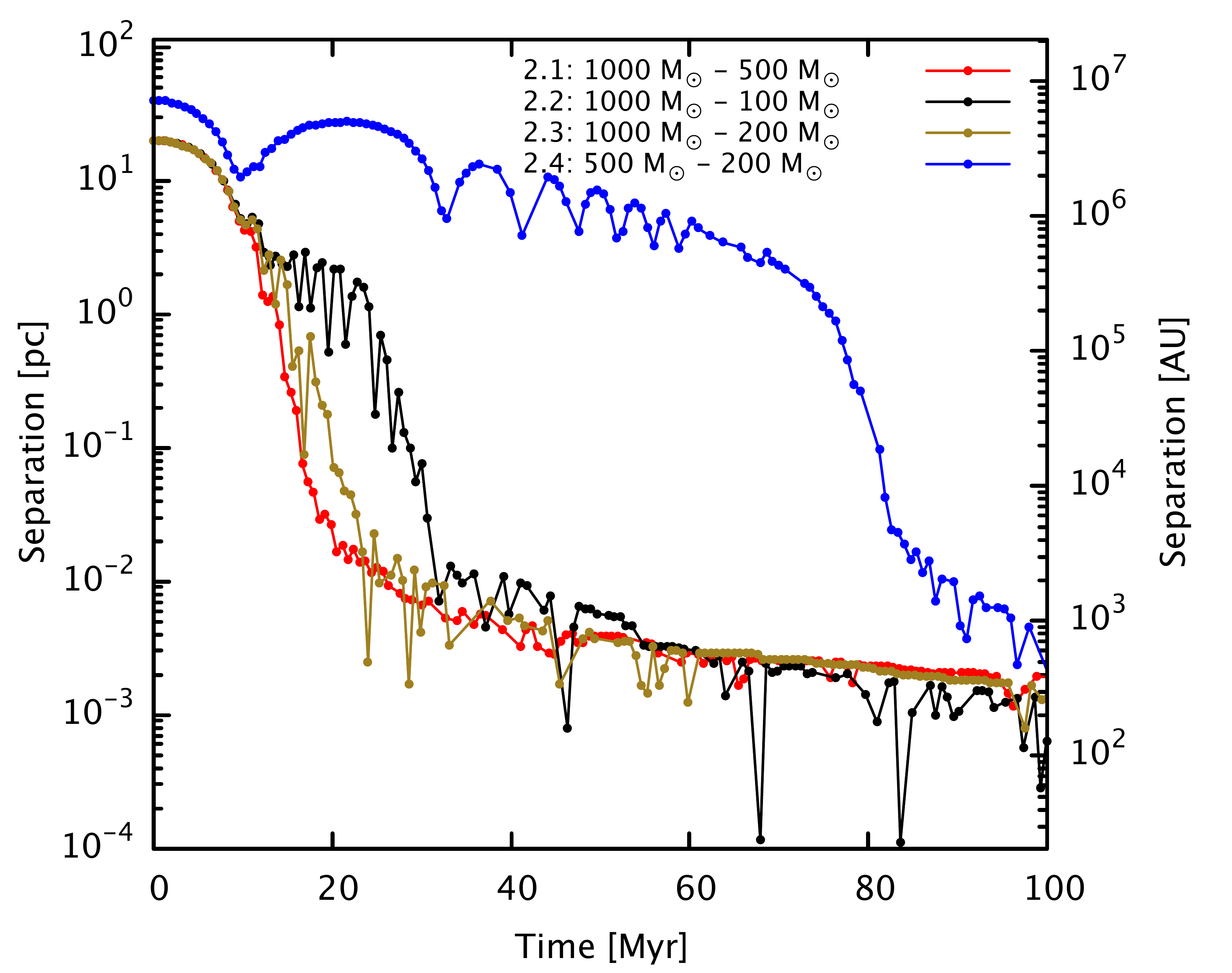}
	  \includegraphics[width=0.48\linewidth]{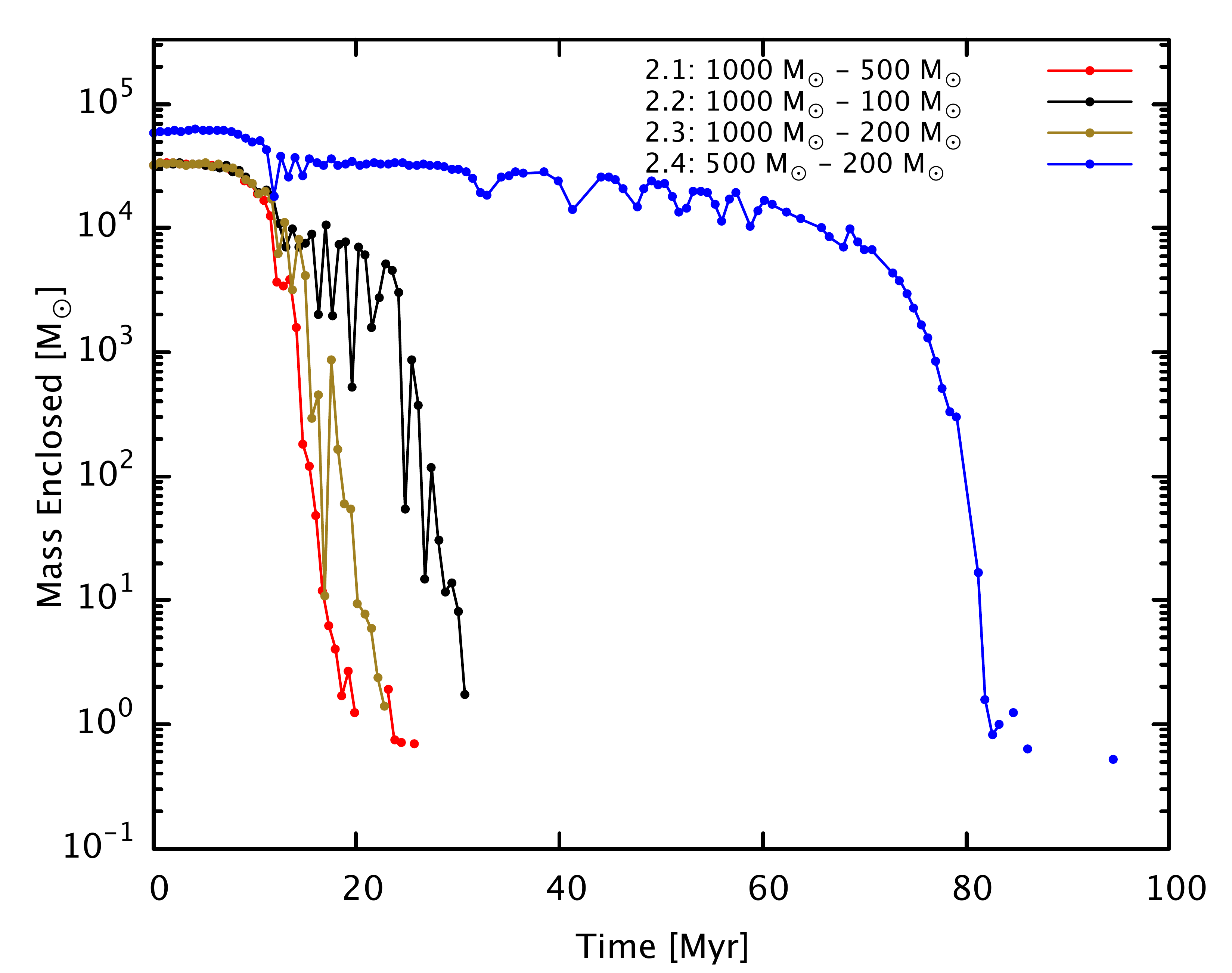}
	\caption{The left panel shows the instantaneous separation between the two IMBHs in the 2.i runs during the first 100 Myr. The red, black, olive and blue lines correspond to runs 2.1, 2.2, 2.3 and 2.4 respectively. The non smooth lines are due to orbital eccentricity of the binary IMBH. The right panel shows the evolution of the mass enclosed between the two IMBHs as functions of time in the four 2.i runs. At the ends of the lines the IMBH binaries have evacuated all of the stars originally contained within their orbits.}
    \label{fig:2x-sep}
\end{figure*}

The left panel in Fig.~\ref{fig:2x-sep} shows the instantaneous separation of the two IMBHs as a function of time for the first 100 Myr. The non-smooth lines in the left panel are due to the orbital eccentricity of the binary IMBH system. For models labelled 2.1 to 2.3, the binary forms within 20-30 Myr. The model with the most massive secondary IMBH (2.1), segregates and finds the primary IMBH within 20 Myr and the model with the least massive secondary IMBH (2.2) takes slightly longer to form a binary system with the primary IMBH. For the model labelled 2.4, there was no central IMBH so the initial separation between the two IMBHs is about 40 pc. For this run, it takes the IMBHs about 100 Myr to form a close binary system.

\begin{table}
\centering
  \caption{This table describes the four runs that contain two IMBHs. The description of the columns is the same as it was for Table \ref{tab:1-imbh-models}. The runs have been labelled with the colours that are used for them in Figs. \ref{fig:2x-sep}, \ref{fig:2x-binary},  \ref{fig:5x-binding-energy} and \ref{fig:den-around-IMBH}.}
  \label{tab:2-imbh-models}
\begin{tabular}{|l|l|l|c|c|c|}
\hline
\textbf{Run} & {\textbf{Time}} & {\textbf{Total Mass}} & \textbf{$\rm IMBH_{1}$} & \textbf{$\rm IMBH_{2}$} & \textbf{$\rm IMBH_{3}$} \\ & [Myr] & [$\rm M_{\odot}$] & [$\rm M_{\odot}$] & [$\rm M_{\odot}$] & [$\rm M_{\odot}$]  \\ \hline
2.1 {\color{red} \textbullet} & 716 & $\rm 8.55 \times 10^{4}$ & 1000 & 500 & 0 \\
2.2  {\color{black} \textbullet} & 914 & $\rm 8.50 \times 10^{4}$ & 1000 & 100 & 0 \\
2.3 {\color{olive} \textbullet} & 825  & $\rm 8.52 \times 10^{4}$ & 1000 & 200 & 0 \\
2.4 {\color{blue} \textbullet} & 745 & $\rm 8.47 \times 10^{4}$ & 0 & 500 & 200 \\ \hline
\end{tabular}
\end{table}

The right panel in Fig.~\ref{fig:2x-sep} shows the evolution of the mass enclosed between the two IMBHs as a function of time in the four 2.i runs. The enclosed mass is calculated by summing up the mass of stars that are enclosed within a sphere centred around the midpoint between the two IMBHs and its radius is given by the length of this midpoint between the two IMBHs. At the end of the lines in the right panel of Fig.~\ref{fig:2x-sep}, the IMBH binaries have evacuated all of the stars originally contained within their orbits by scattering them away \citep{makino2004,berczik2005,merritt2006}.

\begin{figure*}
     \includegraphics[width=0.48\linewidth]{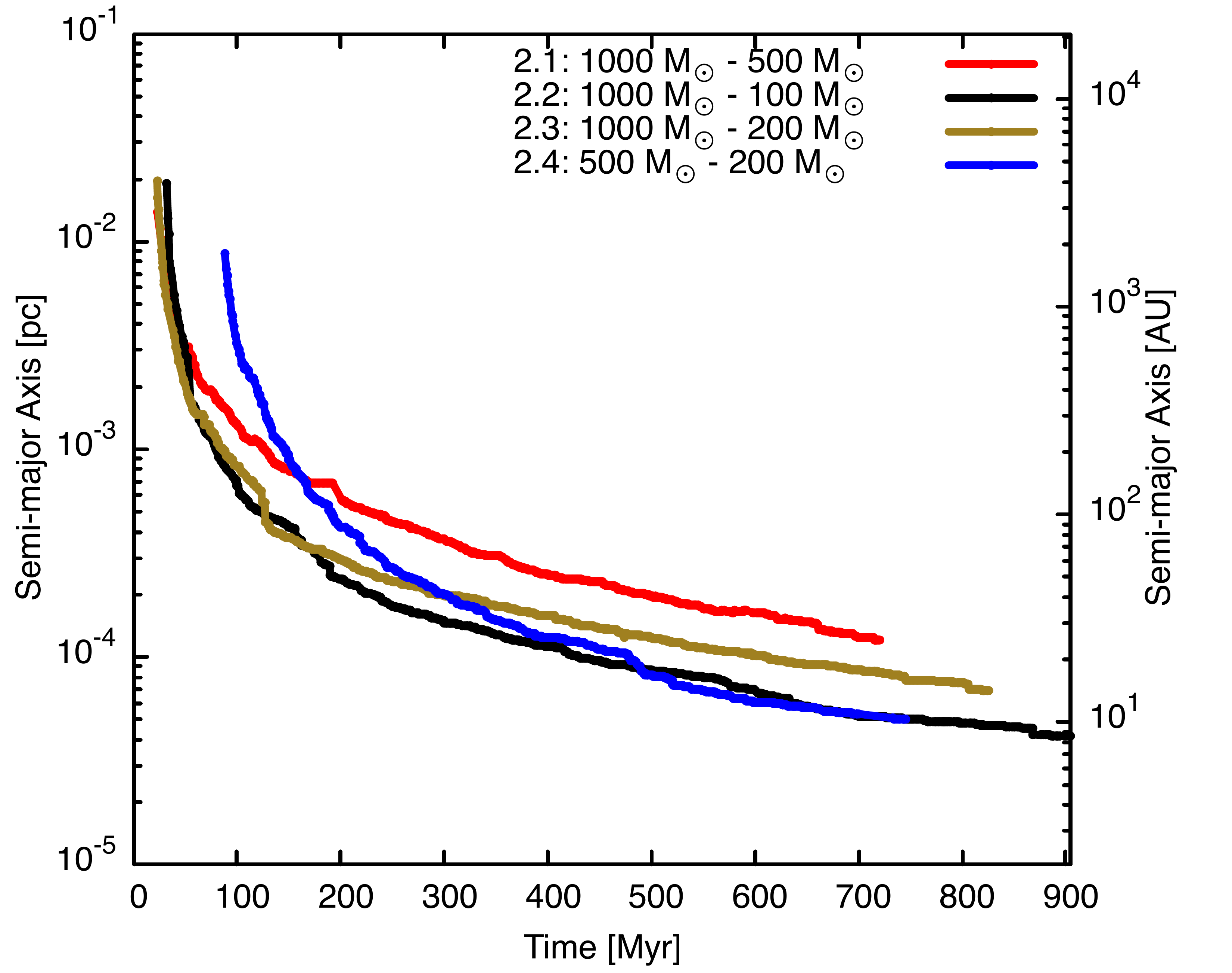}
	 \includegraphics[width=0.48\linewidth]{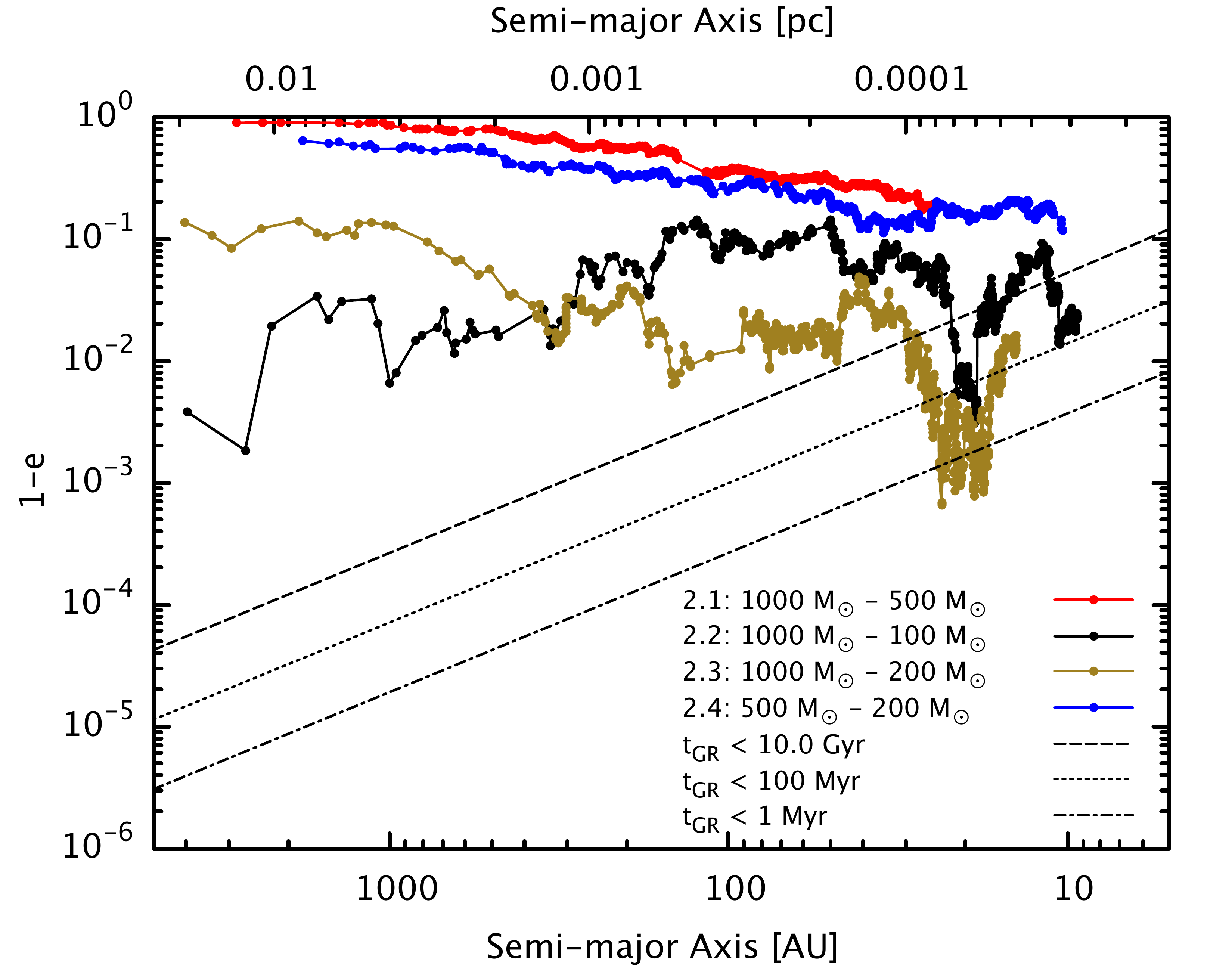}
	\caption{The left panel shows the evolution of the semi-major axis of the binary IMBH following its formation till the end of the run. The right panel shows the semi-major axis vs $1-e$ for the binary in runs 2.i. Lines indicating the time needed for a 1000 $\rm M_{\odot}$ and a 100 $\rm M_{\odot}$ IMBH in a binary to merge via gravitational wave (GW) emission are also shown on the figure. The eccentricity in models 2.2 and 2.3 becomes quite large and the IMBH could merge through GW radiation within few to several hundred Myr respectively.}
    \label{fig:2x-binary}
\end{figure*}

We follow the evolution of the merging clusters for several hundred Myr. The long-term evolution of the semi-major axis and the orbital eccentricity of the binary IMBH in each of the 2.i runs is shown in the left panel of Fig.~\ref{fig:2x-binary}. We find that in all these runs, the binary hardens with time and over the course of a few hundred Myr, the binary semi-major axis decreases from a few thousand AU to a few tens of AU. The binary hardens by scattering away stars in 3-body interactions.  We find that in all the runs, this scattering away of stars by the binary also leads to systematic growth in its orbital eccentricity. The right panel in Fig.~\ref{fig:2x-binary} shows evolution of the semi-major axis and the eccentricity of the binary in the 2.i runs. The evolution of the binary IMBH and the reason for the increase in eccentricity are discussed in Appendix \ref{ross-section}. The merger timescale due to GW emission strongly depends on the orbital eccentricity of a binary. As eccentricity approaches unity, the time needed for a binary to merge due to gravitational radiation emission decreases significantly as it depends on $(1-e{^2})^{\frac{7}{2}}$ \citep{peters1964}.

From the right panel in Fig.~\ref{fig:2x-binary}, it can be seen that the eccentricity for the binary IMBH in runs 2.2 and 2.3 reaches very high values with $1-e$ being less than $10^{-3}$. The merger time due to GW emission for binaries with these eccentricities is less than a few hundred Myr. For runs 2.1 and 2.4, the eccentricity does not reach such high values, hence these binaries will take up to a few Gyr to merge. The main difference between runs 2.2, 2.3 and runs 2.1, 2.4 is the eccentricity at the formation of the binary \citep{nasim2020}. Runs 2.2 and 2.3 form with moderately high eccentricity while the formation eccentricity is lower for runs 2.1 and 2.4. The initial eccentricity is driven by motion of the IMBH during the early phase of its inspiral and determines whether the binary can reach a high eccentricity state within a Gyr in order to effectively merge due to GW radiation. As discussed in Appendix \ref{ross-section}, the increase in binary orbital eccentricity due to binary-single scattering in these runs, which has also been shown in other studies, is most likely a physical effect and does not rely on the assumptions or methods that are used. This includes methods and techniques such as scattering experiments, full $N$-body but low mass resolution and full $N$-body but low number of particles.

\begin{figure}
     \includegraphics[width=1.0\columnwidth]{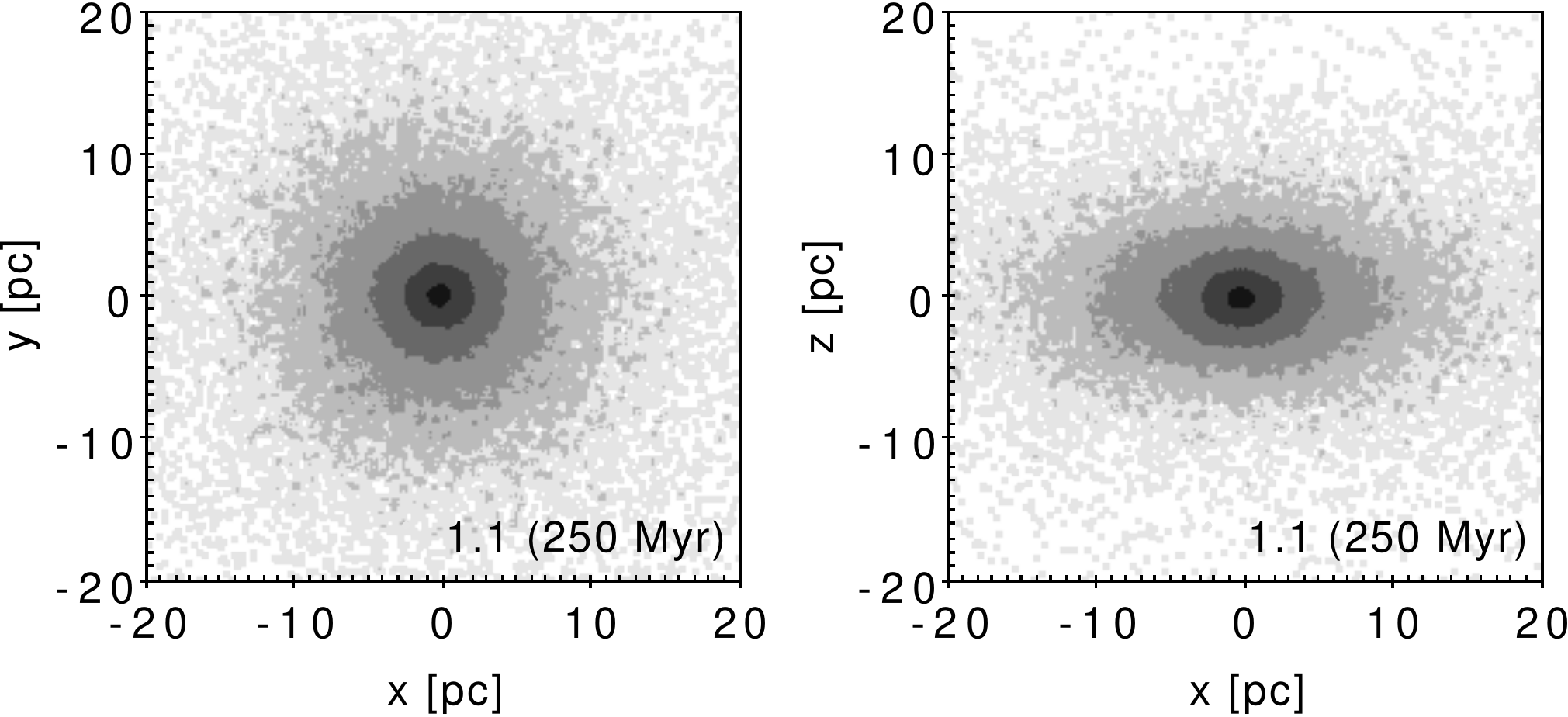}
	 \includegraphics[width=1.0\columnwidth]{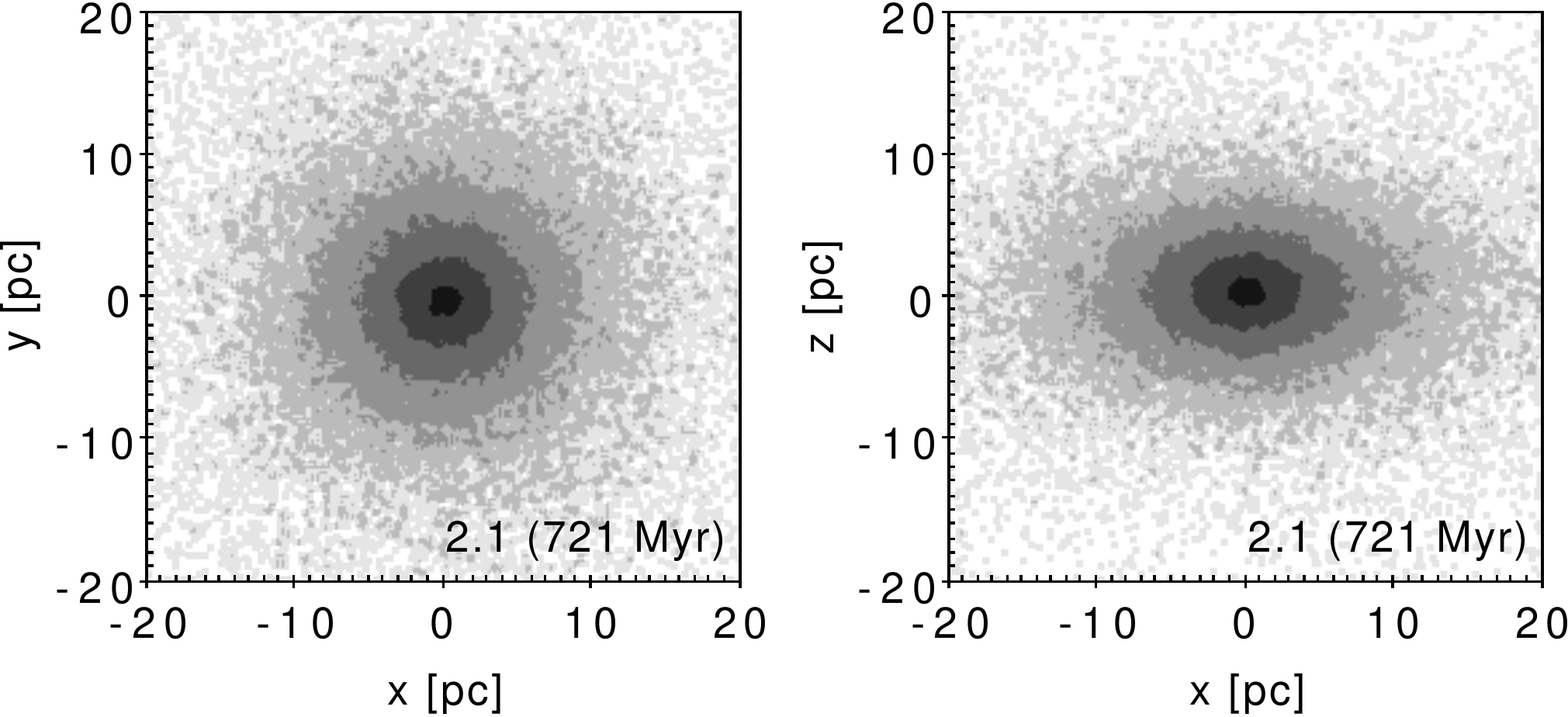}
	 \includegraphics[width=1.0\columnwidth]{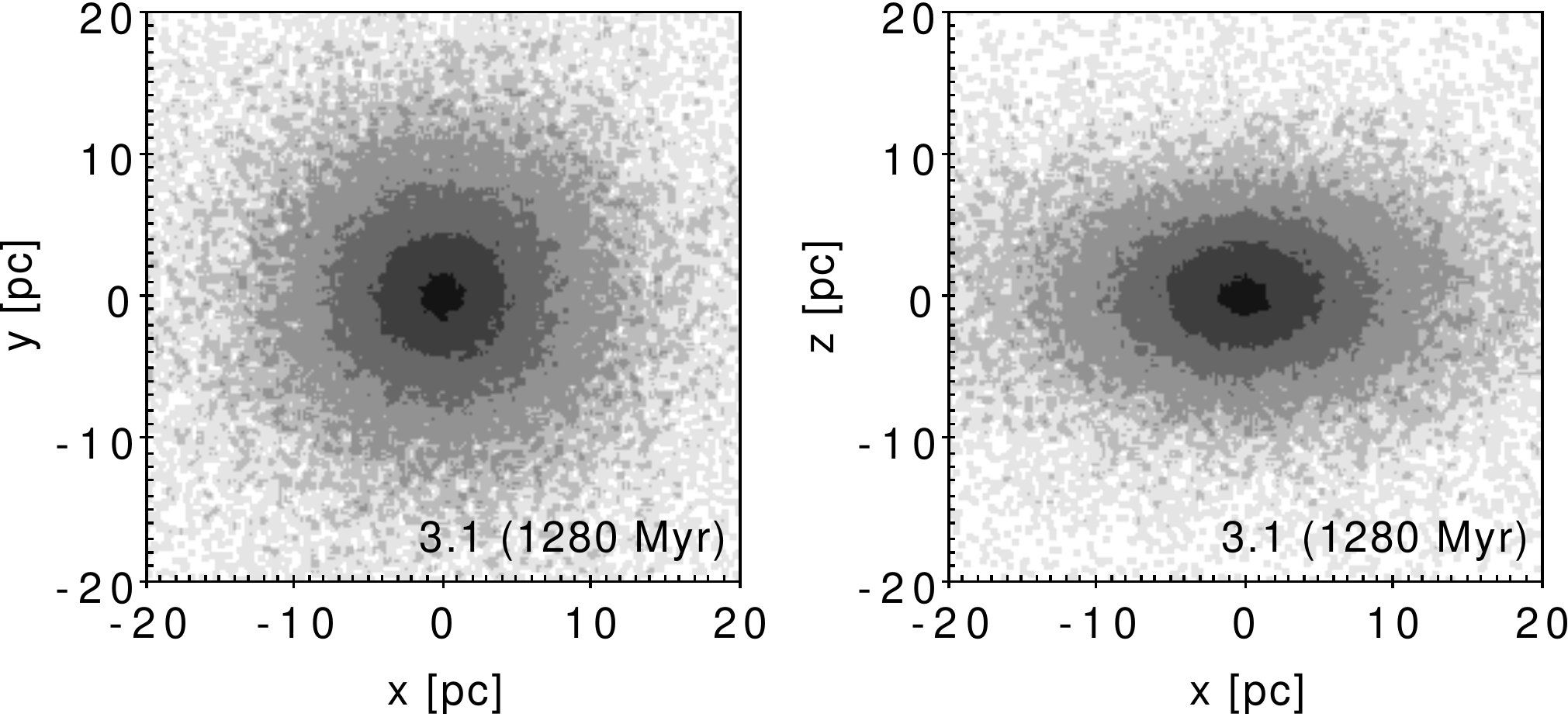}
	\caption{The three rows of figure show the density shaded $X-Y$ (left) and $X-Z$ (right) particle plots for runs 1.1 (top panel), 2.1 (middle panel) and 3.1 (lower panel) using the simulation snapshots at 250, 721 and 1280 Myr, respectively. The origin in these plots is the density centre of the merged cluster.}
    \label{fig:ellipticiy}
\end{figure}

The intial setup of the merger between the clusters can impart a residual rotation on the final cluster. We find that the IMBH binary orbital plane is co-planar with respect to the merged cluster rotation plane for our runs. As discussed by \citet{iwasawa2011}, this can result in preferential ejection of stars on prograde orbits which can leave more stars on a retrograde orbit around the IMBH binary leading to transfer of angular momentum from the secondary to field stars resulting in an increase in eccentricity. This process has also been observed in the evolution of massive BH binary perturbed by a massive star cluster in a galactic centre \citep{sedda2019c}. Additionally, the setup of the merging clusters leads to flattening of the merged cluster in the $Z$ direction. This ellipticity of the merged cluster can also further drive the hardening and eccentricity of the binary IMBH by rapidly refilling the loss cone \citep{khan2013,khan2018,khan2020}. Fig. \ref{fig:ellipticiy} shows the density shaded $X-Y$ and $X-Z$ plane plots of the merged cluster for runs containing 1, 2 and 3 IMBH at different snapshot times. From the plots it can be clearly seen that the cluster is flattened in the $Z$ direction and the typical values of $c / a$ axis ratio range between $\sim$ 0.5 to 0.8.

\section{Simulations containing three IMBHs (3.i runs)}\label{sec4:3i-runs}

We simulated three runs in which each of the three stellar clusters in our original set-up (Section \ref{sec2:model-description}) initially had an IMBH at its centre. The masses of the IMBHs in the three merging stellar clusters were 1000, 500 and 200\,$\msun$. The configuration of the IMBHs in the runs and the time for which they were run are given in Table \ref{tab:3-imbh-models}.

In run 3.1, the central star cluster with 50\,000 particles hosted a 1000 $\msun$ IMBH, and the infalling clusters with 30\,000 and 15\,000 particles contained 500 and 200 $\rm M_{\odot}$ IMBHs respectively. In run 3.2, the initial position of the 1000 $\msun$ IMBH was swapped with the 500 $\msun$ IMBH.  In run 3.3, the central star cluster with 50\,000 particles contains the 200 $\rm M_{\odot}$ IMBH while the least massive stellar cluster with 15\,000 particles hosts the 1000 $\rm M_{\odot}$ IMBH. 

\begin{table}
\centering
  \caption{Table showing the 3 simulated merging stellar cluster models that contain 3 IMBH. The description of columns is as that in Table \ref{tab:1-imbh-models}. The runs have been labelled with the colours that are used for them in Figs. \ref{fig:3x-inner-bin}, \ref{fig:5x-binding-energy} and \ref{fig:den-around-IMBH}.}
  \label{tab:3-imbh-models}
\begin{tabular}{|l|l|l|c|c|c|}
\hline
\textbf{Run} & {\textbf{Time}} & {\textbf{Total Mass}} & \textbf{$\rm IMBH_{1}$} & \textbf{$\rm IMBH_{2}$} & \textbf{$\rm IMBH_{3}$} \\ & [Myr] & [$\rm M_{\odot}$] & [$\rm M_{\odot}$] & [$\rm M_{\odot}$] & [$\rm M_{\odot}$]  \\ \hline
3.1 {\color{cyan} \textbullet} & 1511 & $\rm 8.57 \times 10^{4}$ & 1000 & 500 & 200 \\
3.2 {\color{light-violet} \textbullet} & 832 & $\rm 8.57 \times 10^{4}$ & 500 & 1000 & 200 \\
3.3 {\color{violet} \textbullet} & 939 & $\rm 8.57 \times 10^{4}$ & 200 & 500 & 1000 \\ \hline
\end{tabular}
 \end{table}

\begin{figure*}
	 \includegraphics[width=0.32\linewidth]{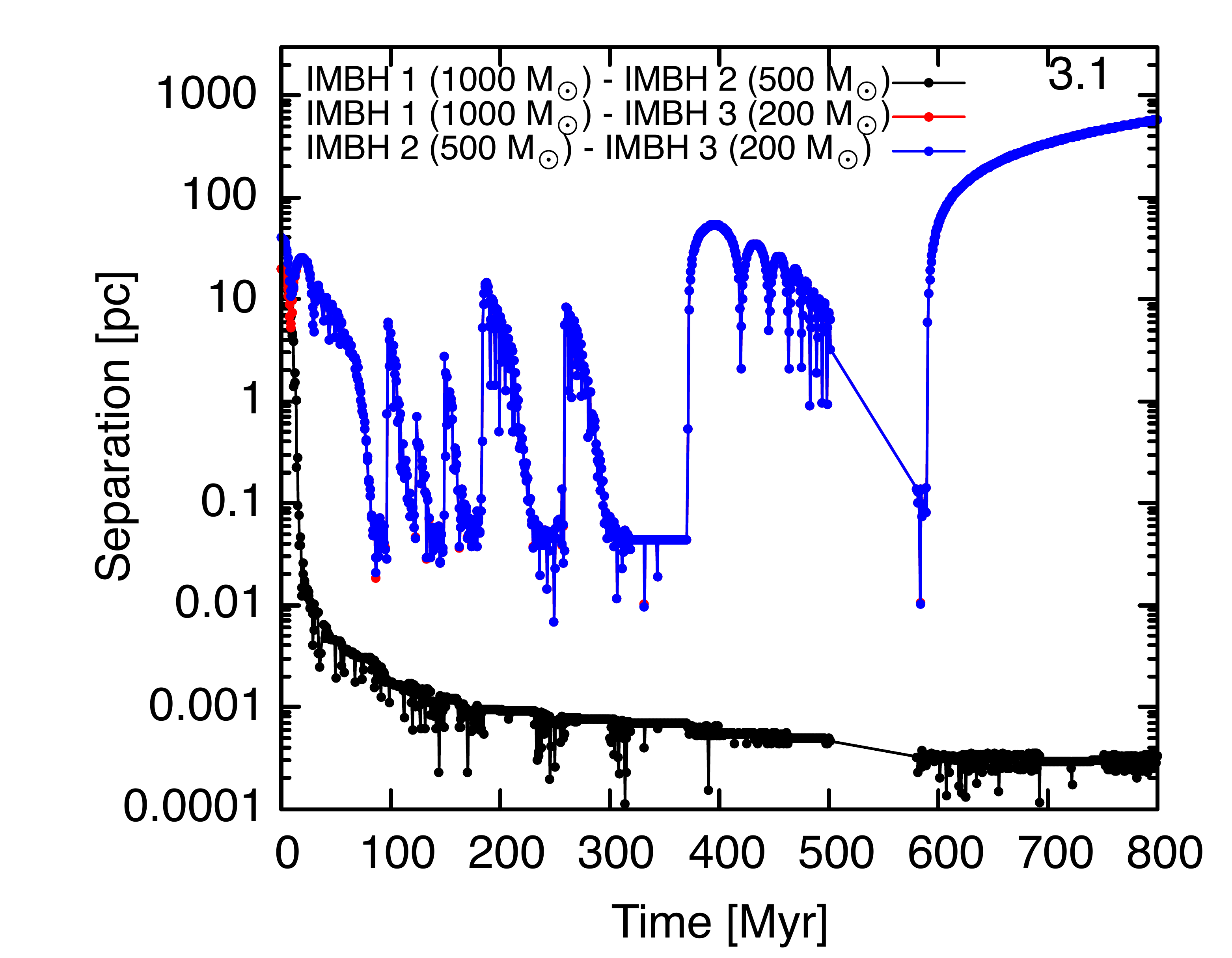}
	  \includegraphics[width=0.32\linewidth]{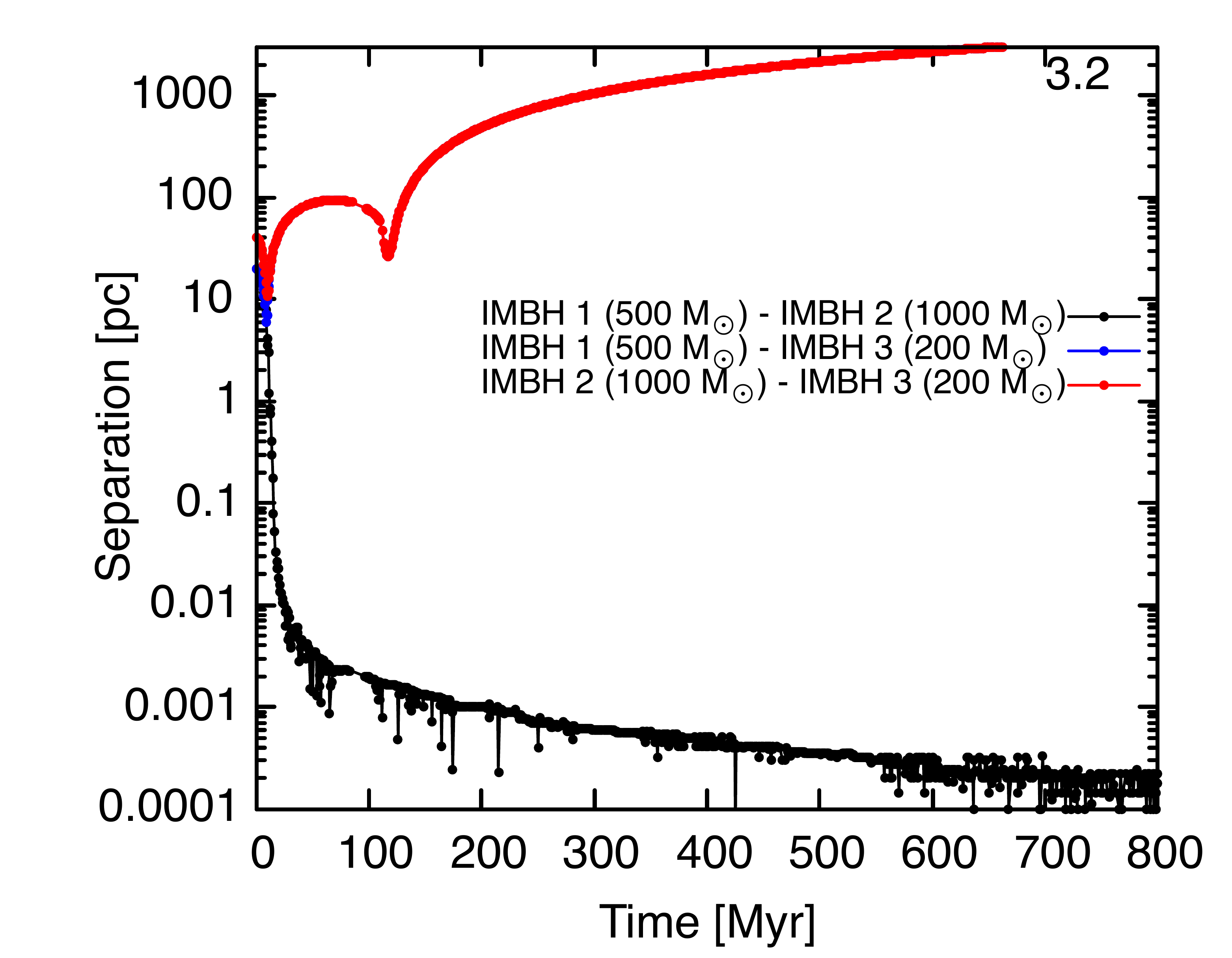}
	  \includegraphics[width=0.32\linewidth]{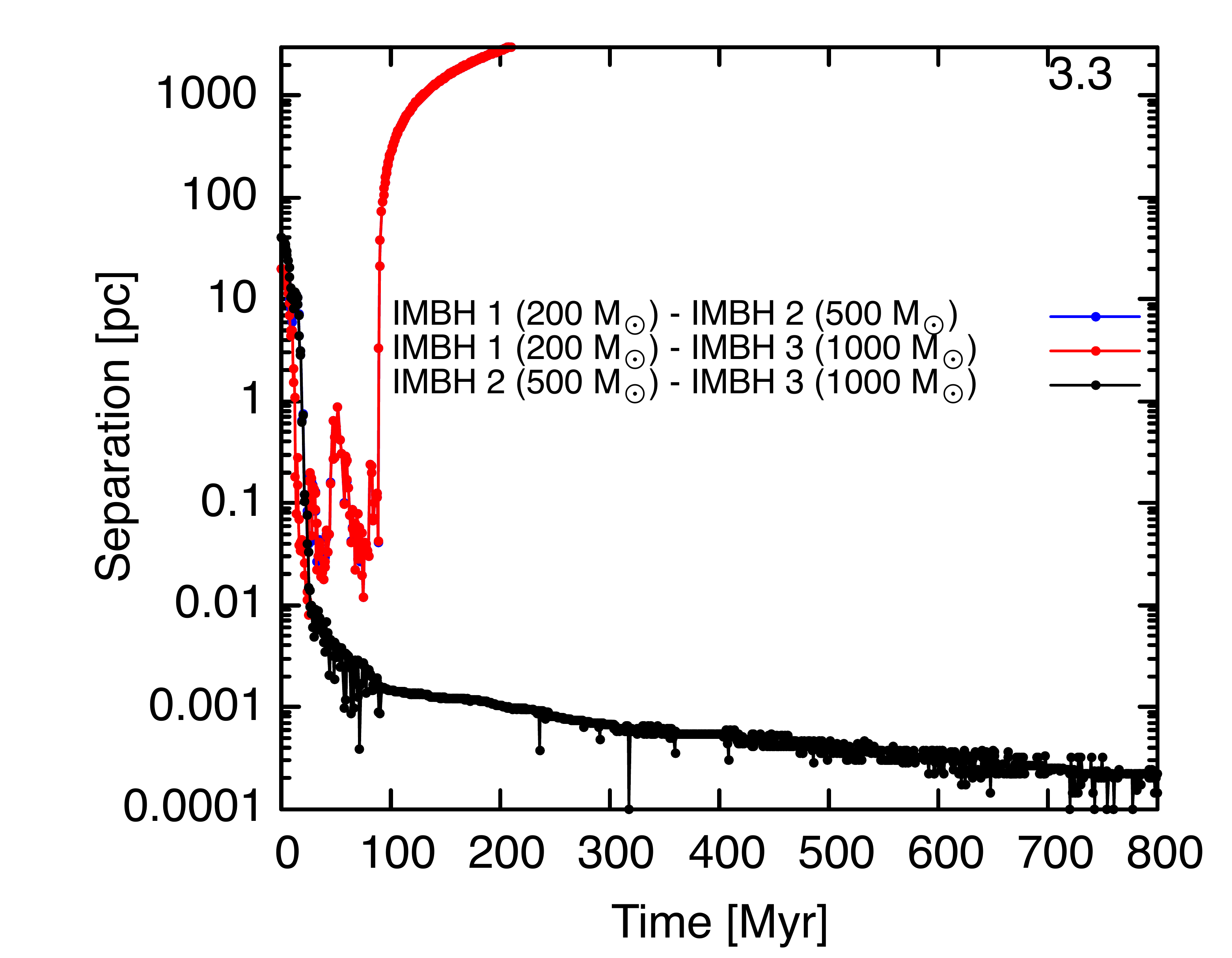}
	\caption{The three panels show the instantaneous separation between the three IMBHs in the 3.i runs. The non-smooth lines are due to the orbital eccentricity of the binary IMBH system. In all the 3.i runs, we form an IMBH binary consisting of the 1000 $\msun$ and 500 $\msun$ IMBH. For run 3.1, the precise data for the IMBHs between 500 to about 580 Myr was not available as they were part of  a chain interaction in \textsc{nbody6++GPU}.}
    \label{fig:3x-sep}
\end{figure*}

In all of the three runs, the 1000 and 500\,$\rm M_{\odot}$ IMBHs segregate to the centre of the merged cluster to form a binary IMBH. We show the separation between the three IMBHs in Fig.~\ref{fig:3x-sep}.

Similar to the 2.i runs, the two IMBHs end up forming a binary on timescales of a few tens of Myr. In all these runs, the two most massive IMBHs end up in a binary at the centre of the merged cluster. In runs 3.1 and 3.3, the 200\,$\msun$ IMBH also sinks to the centre and reaches a separation of $10^{-2}$\,pc from the binary IMBH. In both these runs, the 200\,$\msun$ IMBH is ejected from the merging clusters because of a strong encounter with the binary IMBH. 

\subsection{Run 3.1: Multiple strong binary-single IMBH scatterings}\label{subsec:strong3}

\begin{figure}
	 \includegraphics[width=\columnwidth]{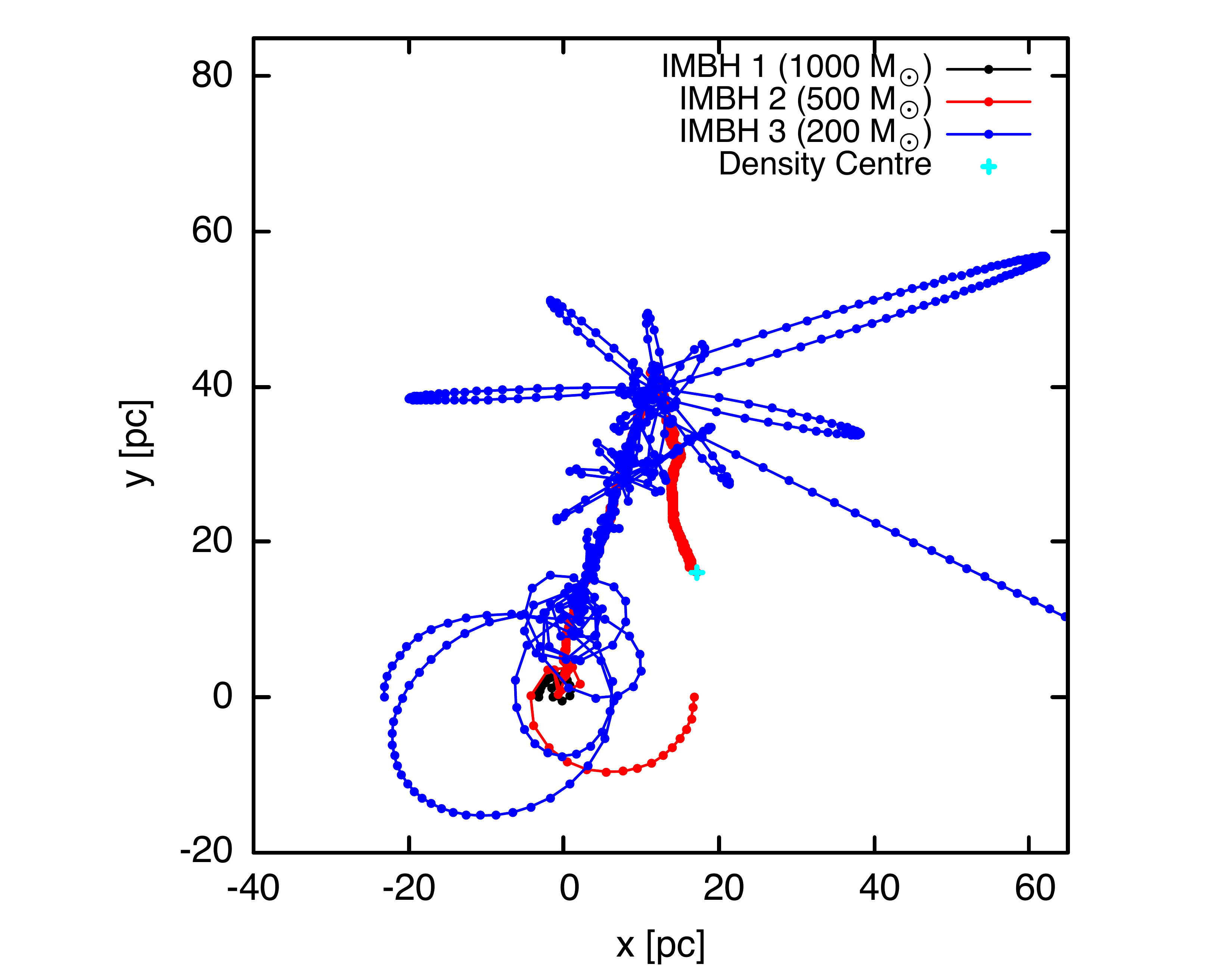}
	\caption{$X-Y$ position of the 3 IMBHs in run 3.1. The 1000 and 500 $\rm M_{\odot}$ IMBHs quickly form a close binary system within 25 Myr. The $200\,\rm M_{\odot}$ IMBH (shown in blue) slowly segregates to the cluster centre and a triple IMBH system forms. Subsequently the 200 $\rm M_{\odot}$ IMBH is scattered away from the inner binary IMBH repeatedly until it is finally ejected. The cyan cross indicates the $X-Y$ position of the centre of density at the last snapshot.}
    \label{fig:31-xy}
\end{figure}

\begin{figure}
	 \includegraphics[width=\columnwidth]{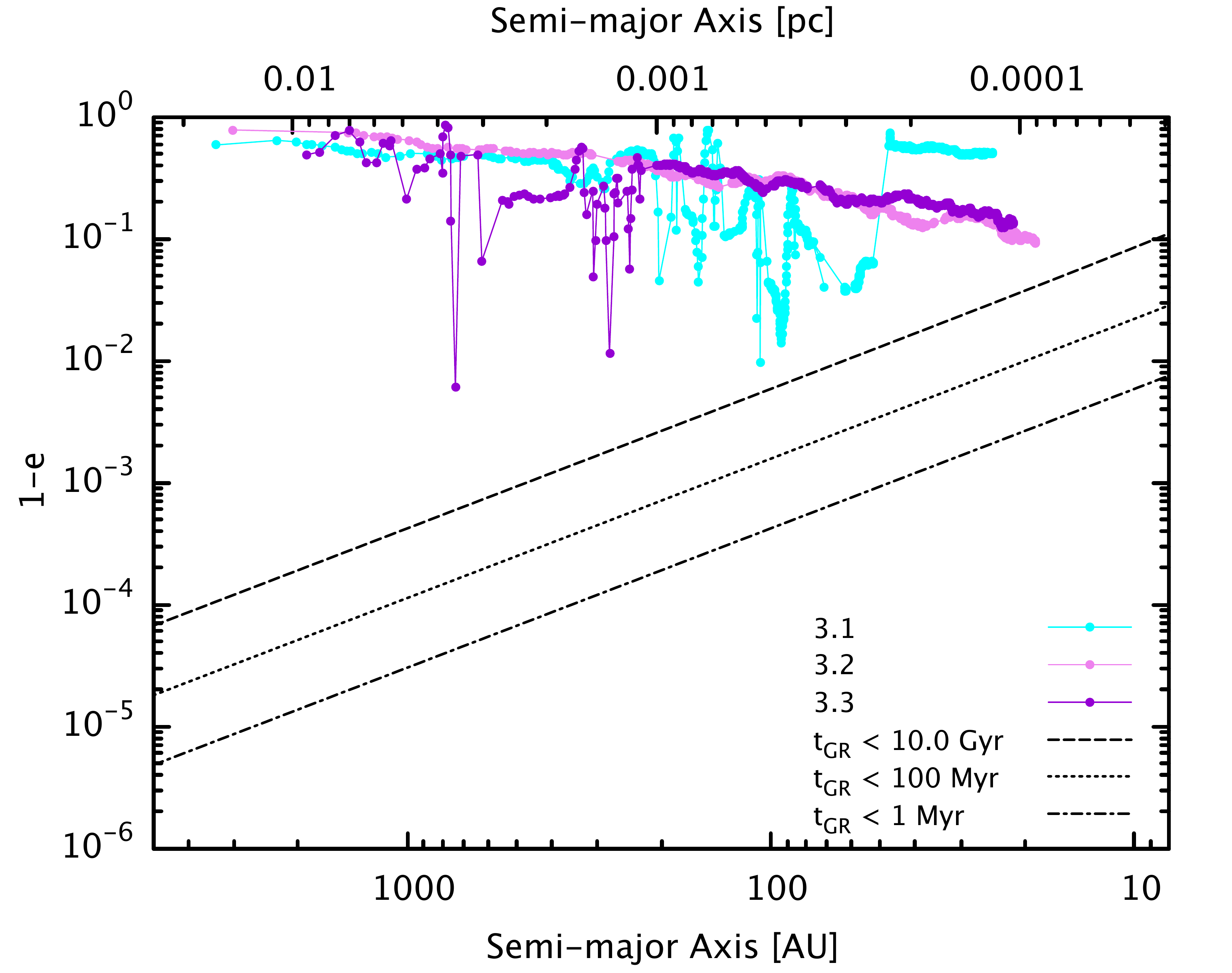}
	\caption{Semi-major axis vs 1-eccentricity of the inner binary IMBH containing the 1000 $\rm M_{\odot}$ and 500 $\rm M_{\odot}$ IMBH in 3.i runs. The cyan line is for run 3.1, the violet line is for run 3.2 and the dark violet line is for run 3.3.}
    \label{fig:3x-inner-bin}
\end{figure}

In run 3.1, we find that the third IMBH undergoes multiple strong encounters with the binary IMBH. The $X-Y$ position plot of the three IMBHs in run 3.1 is shown in Fig.~\ref{fig:31-xy}. The binary IMBH forms with an eccentricity of 0.4. Initially, the evolution of the binary is similar to run 2.1 and we find that its eccentricity gradually increases as the semi-major axis decreases. However, frequent interactions with the third IMBH significantly modify the eccentricity evolution, as seen in Fig.~\ref{fig:3x-inner-bin}. When the binary has a semi-major axis of around 200 AU, we see large changes in the eccentricity of the binary which are caused by interactions with the 200 $\msun$ IMBH. The interaction which leads to the ejection of the $200\,\msun$ IMBH takes place at about 580 Myr (see Fig.~\ref{fig:3x-sep}, leftmost panel). The binary remains intact and bound to the cluster, however, its trajectory within the cluster changes. The strong scattering that ejects the $200\,\msun$ IMBH also causes a sharp decrease in eccentricity of the binary IMBH. The semi-major axis of the binary IMBH is 50 AU at this time. Subsequently, the eccentricity gradually increases again and evolves similarly to the 2.1 run. However, the hardening of the binary is slower compared to run 2.1 because frequent interactions between the three IMBH and surrounding stars after 100 Myr results in the ejection of a significant number of stars from the central part of the cluster. Also the recoil from the ejection of the third IMBH (see Section \ref{sec4:binsin-interaction}) puts the binary in a part of the cluster where the background stellar density and velocity dispersion is lower. The density of stars inside a box of 1 pc around the centre of mass of the binary IMBH at 582\,Myr (shortly after the ejection) was $\rm 340 \, \msun \, pc^{-3}$ for run 3.1, compared to $\rm 1126 \, \msun \, pc^{-3}$ in run 2.1 (see left panel in Fig. \ref{fig:den-around-IMBH}). The slower hardening of the binary in run 3.1 after 580 Myr can also be seen in Fig.
\ref{fig:5x-binding-energy}, which shows the evolution of the binding energy of the binary IMBHs in runs 2.i and 3.i. This slower hardening of the binary IMBH increases the time needed for it to merge due to GW radiation. Extrapolating the evolution of the gravitational wave merger time using the semi-major axis and eccentricity values for the IMBH binary from snapshots after 600 Myr, we estimate a merger time of around 6000 Myr. We find that after 1500 Myr of evolution, the IMBH binary is in close proximity ($\lesssim 0.2 \ \rm pc$) to the cluster density centre.

\subsection{Run 3.2: Tidal stripping of third IMBH prevents triple formation}\label{subsec:imbh-stripping}

In run 3.2, as the merging clusters spiral in together, the 200 $\msun$ IMBH is put on a wide orbit along with stars that are stripped away from its host cluster. The 200 $\msun$ IMBH never comes close to the inner binary and hence does not affect its evolution. The semi-major axis and eccentricity evolve in a similar way to run 2.1, since the eccentricity of the binary at formation time is similarly low (0.2 for run 3.2 compared to 0.1 for run 2.1).

\subsection{Run 3.3: Early ejection of the $200\,\msun$ IMBH}

In run 3.3, the $200\,\msun$ IMBH is ejected following an interaction with the binary IMBH at around 90 Myr. At this time, the semi-major axis of the binary is about 200 AU ($\rm \sim 0.001$ pc;  see Fig.~\ref{fig:3x-sep}). Prior to the ejection, the eccentricity of the binary fluctuates due to dynamical interactions with the 200 $\msun$ IMBH. Following this interaction, the eccentricity and semi-major axis of the binary evolve in a similar way to run 2.1. Unlike run 3.1, the interaction that ejects the third IMBH occurs earlier in the evolution of the system when the binary semi-major axis is large. This leads to a lower recoil velocity for the binary and it remains in the centre of the merging cluster where the background stellar density is high.

\subsection{Summary of simulations containing three IMBHs}

In summary, the 3.i runs show diverse outcomes. In run 3.1, the third IMBH stays within the merged cluster for a few hundred Myr before it is ejected in a binary-single scattering. In run 3.2, the lowest mass IMBH ends up on a wide orbit and never dynamically interacts with binary IMBH. In run 3.3, the binary IMBH and the third IMBH have a strong encounter that leads to the ejection of third IMBH within a 100 Myr from the start of the simulation; again, it does not affect the subsequent evolution of the binary.

\begin{figure*}
     \includegraphics[width=0.48\linewidth]{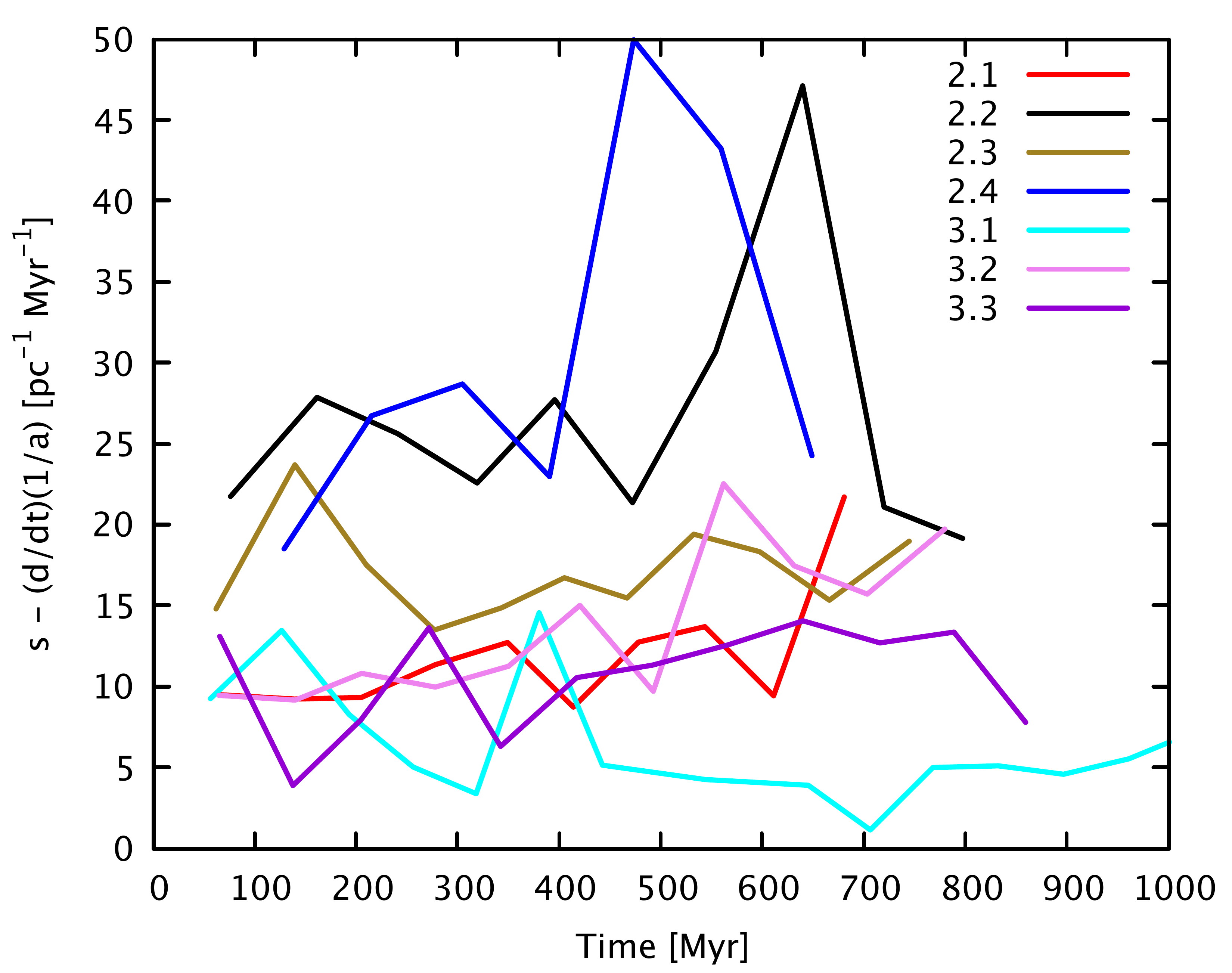}
	 \includegraphics[width=0.48\linewidth]{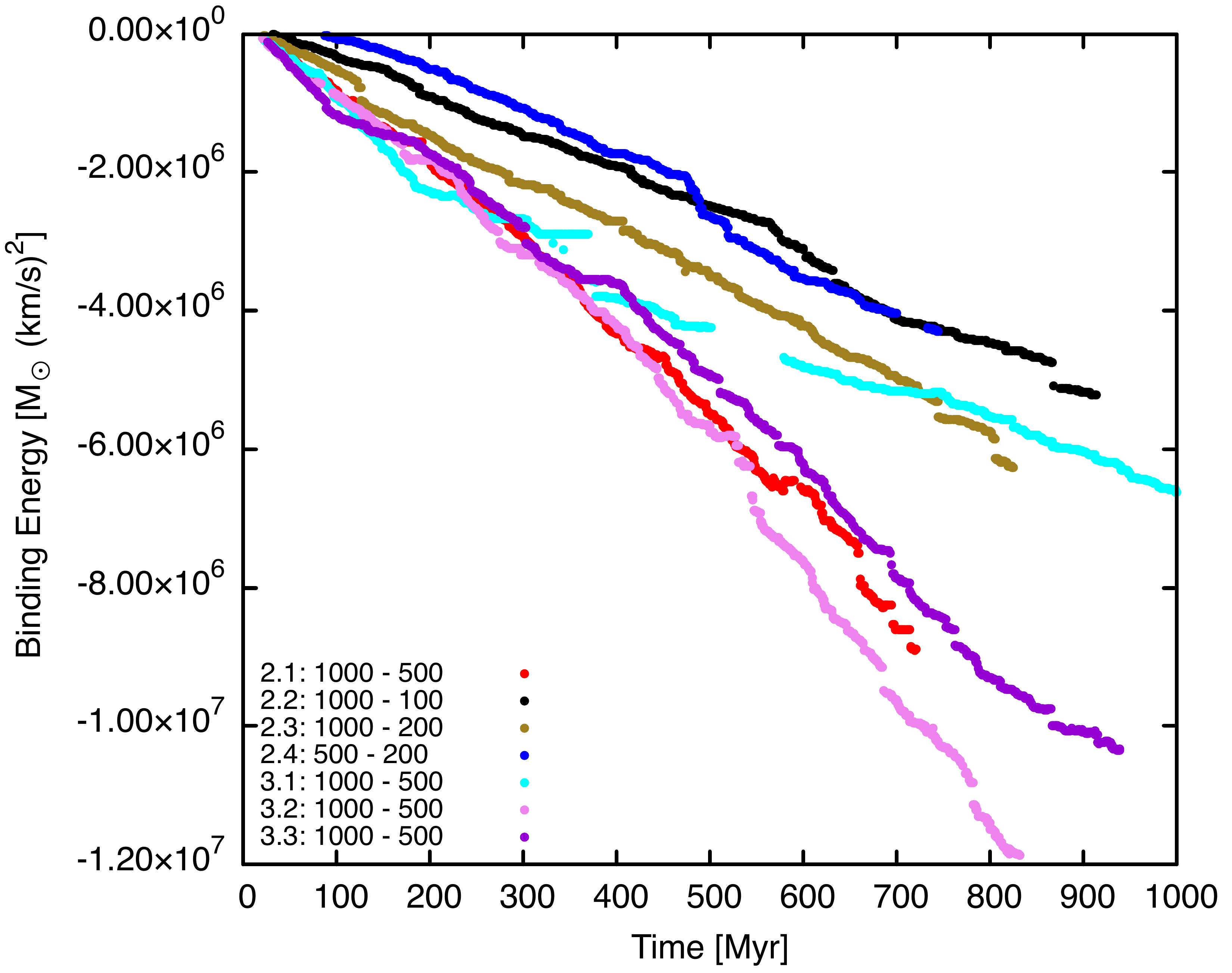}
	\caption{The left panel shows the evolution of the time dependent hardening rate (Equation \ref{eq:hardening})  of the binary of the binary IMBH in the 2.i and 3.i runs. The right panel shows the evolution of the binding energy of the binary IMBH.}
    \label{fig:5x-binding-energy}
\end{figure*}

Similarity in the hardening of the binary IMBH in runs 2.1, 3.2 and 3.3 can be seen in  Fig.~\ref{fig:5x-binding-energy}. The binding energy of the IMBH binaries (1000 $\msun$ - 500 $\msun$) in all three runs increases at a nearly constant rate of $\rm 1.1 \times 10^{4} \, M_{\odot} \, \rm km^2 \, s^{-2} \, Myr^{-1}$. 
In comparison, for run 3.1, the binary hardens at a slower average rate of $\rm 7.6 \times 10^{3} \, M_{\odot} \, \rm km^2 \, s^{-2} \, Myr^{-1}$. As discussed in Section \ref{subsec:strong3}, the binary IMBH strongly interacts with the 200 $\msun$ IMBH during the first 580 Myr of its evolution this results in the ejection of stars from the inner regions of the merged cluster. Eventually, the 200 $\msun$ IMBH is ejected and the binary hardening rate is slowed down due to a lower background stellar density.

We have also calculated the the time-dependent binary hardening rate \citep{quinlan1996}:

\begin{equation}
s \equiv \frac{d}{d t}\left(\frac{1}{a}\right)
\label{eq:hardening}
\end{equation}

by calculating the gradient of the $a^{-1}(t)$ evolution over fixed intervals from the time of the binary formation to the end of the simulations (similar to \citet{gualandris2012,sedda-gualandris2018}). As previously discussed, the hardening of the IMBH binary is higher for the lower mass ratio binaries in runs 2.2 and 2.4. The density of stars around the IMBH binary is also higher in these runs. For runs 2.1, 3.2 and 3.3, the hardening rate is roughly constant. For run 3.1, the presence of the third IMBH in the cluster slows down the hardening rate of the IMBH binary. However, as the model evolves beyond 800 Myr, the hardening rate of the IMBH binary is beginning to increase as it settles back into the cluster centre.

\subsection{Cluster density evolution and stellar properties}\label{stellar-properties}

\begin{figure}
	 \includegraphics[width=\columnwidth]{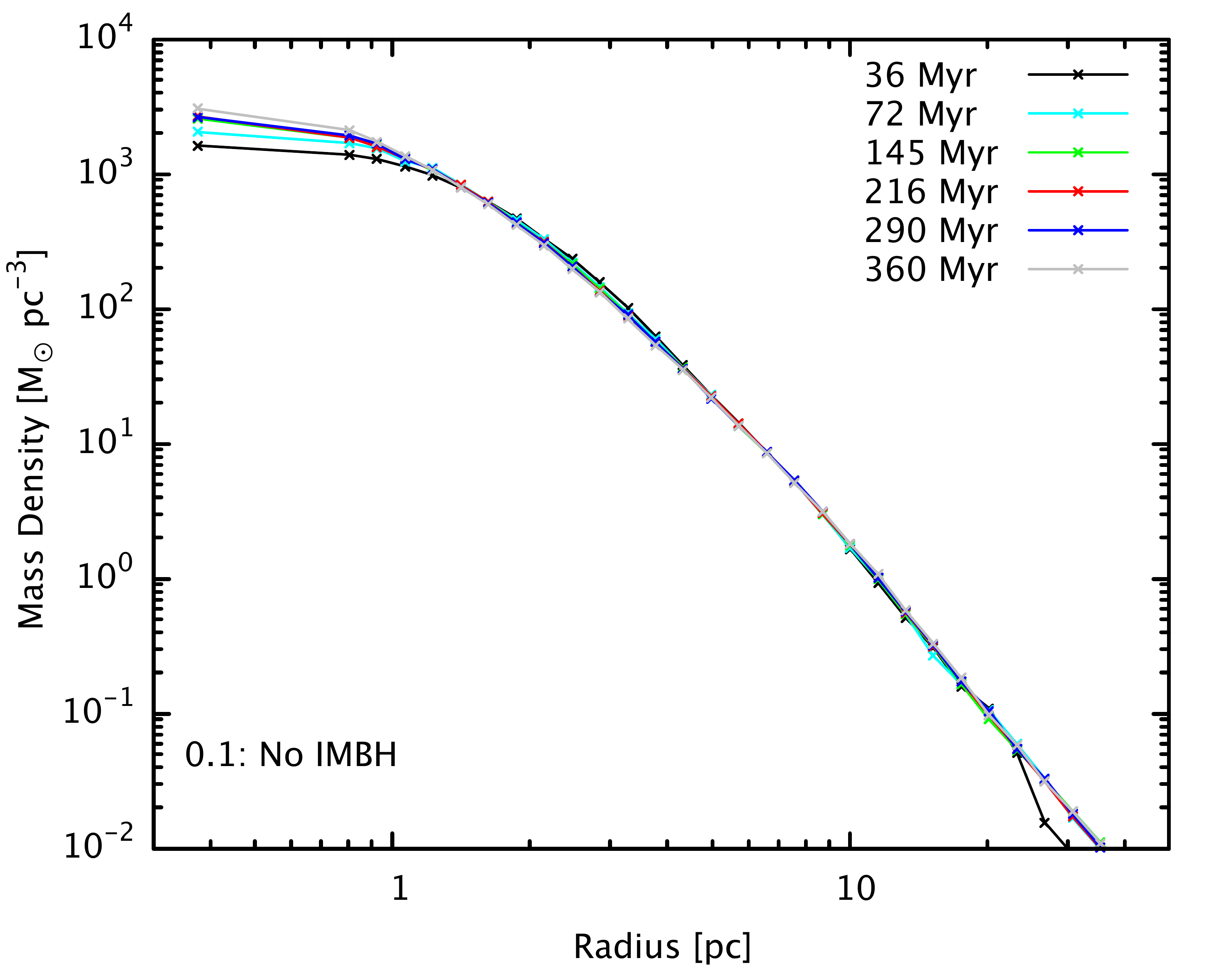}
	  \includegraphics[width=\columnwidth]{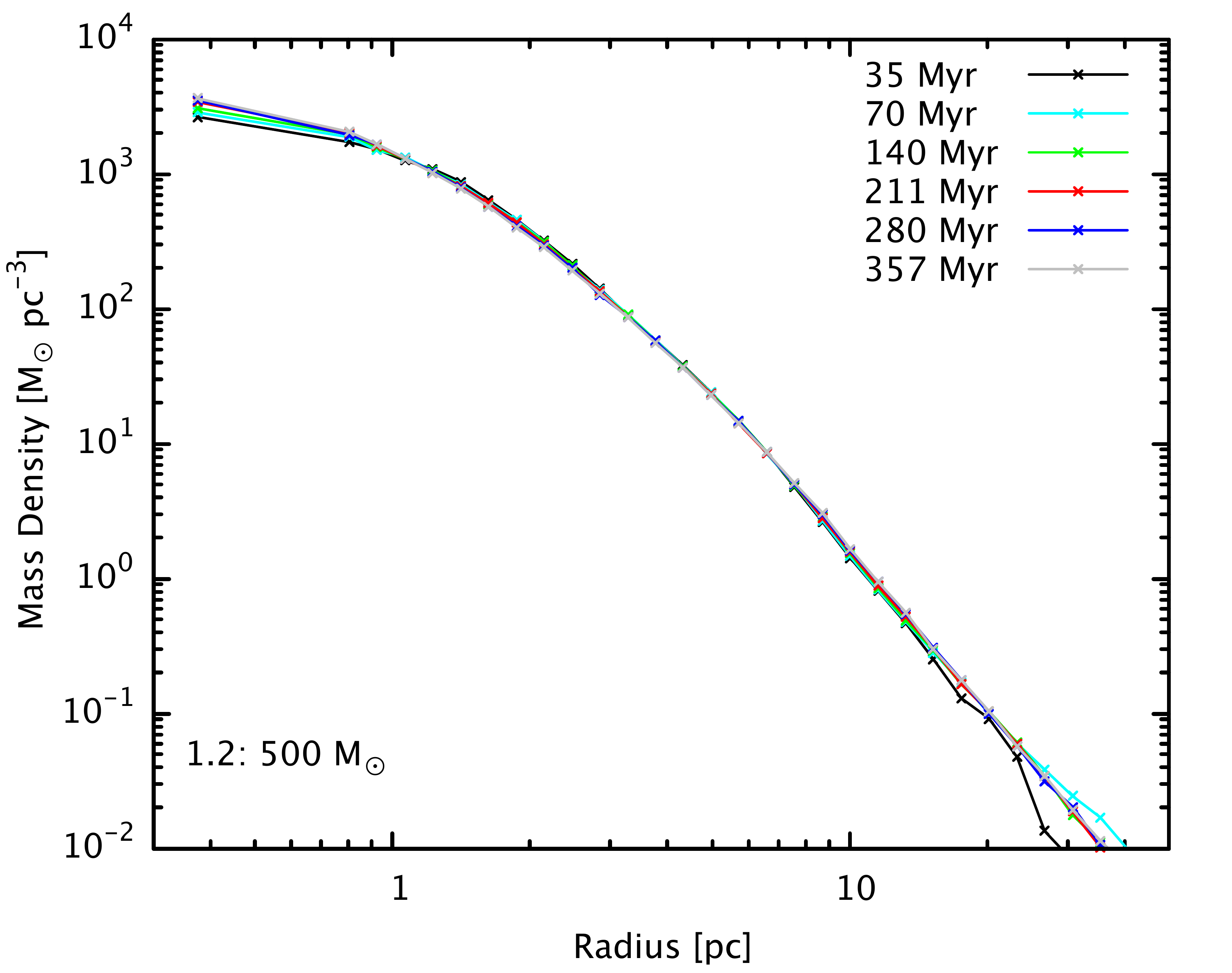}
	\caption{The two panels show the evolution of the radial density profile of the merged cluster for runs 0.1 (top) and 1.2 (bottom).}
    \label{fig:imbh-density-0-1}
\end{figure}

Using the runs described in previous sections, we have also checked how the cluster density structure evolves and what
influence the presence of an IMBH or IMBH binary has on the cluster density profile. It was shown in Fig. \ref{fig:1imbh-density} that the density profile of the the run with no IMBH and the runs with 1 IMBH were similar after a few hundred Myr of cluster evolution. In the two panels of Fig. \ref{fig:imbh-density-0-1}, we show the evolution of the cluster density profile for runs 0.1 (top panel) and 1.2 (bottom panel) using simulation snapshots. For each snapshot, the position of the density-weighted centre was found using the procedure described in \citet{devita2018}. In both these runs, we find that the central density of the cluster increases by a factor of 2-3 at around 360 Myr of evolution. The central density of a merged cluster is expected to increase with time as it dynamically evolves \citep{AMB2019}. The presence of an IMBH also leads to a slightly steeper inner density profile. A similar evolution for the density profile is also observed for runs 1.1 and 1.3.

\begin{figure*}
     \includegraphics[width=0.48\linewidth]{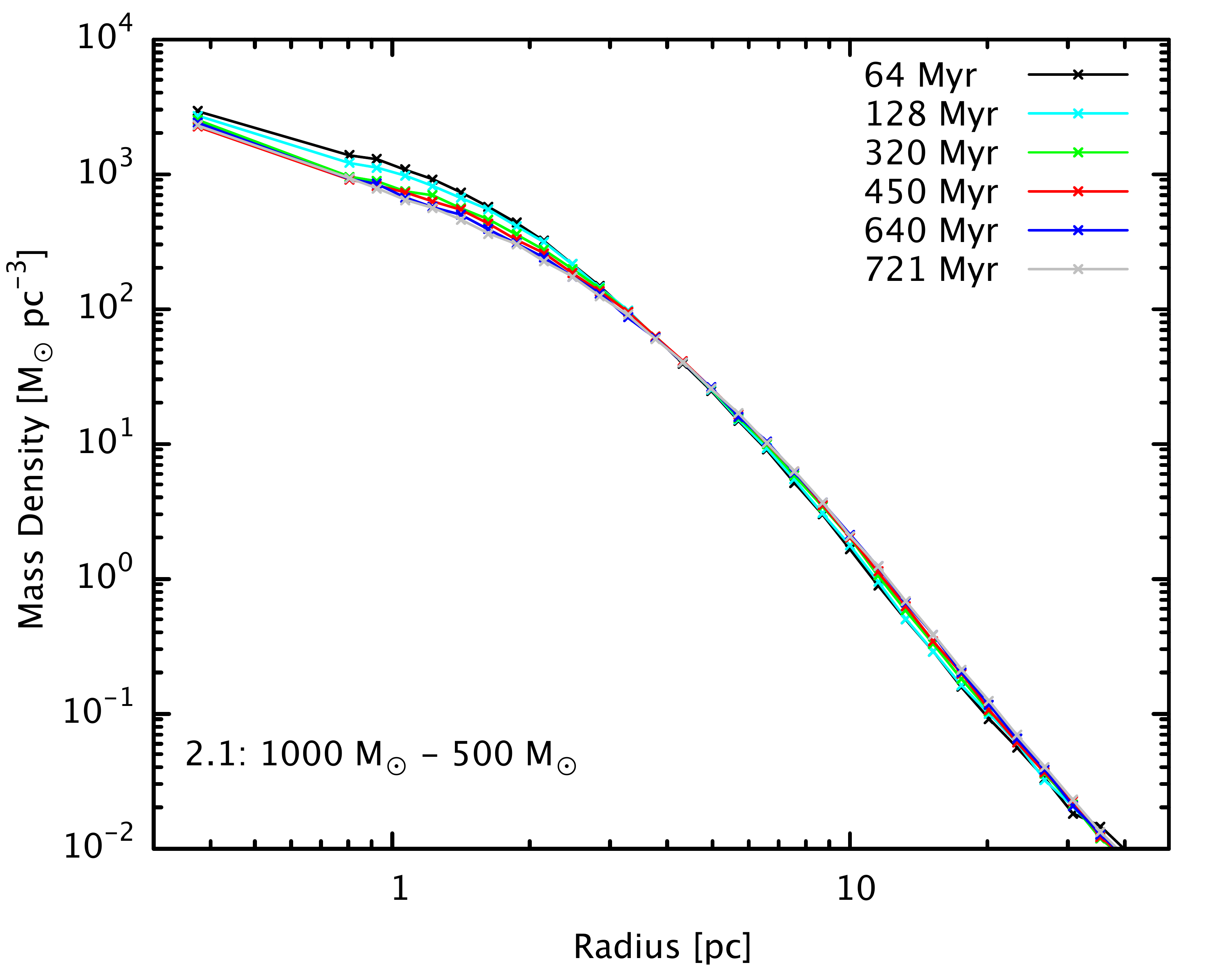}
	 \includegraphics[width=0.48\linewidth]{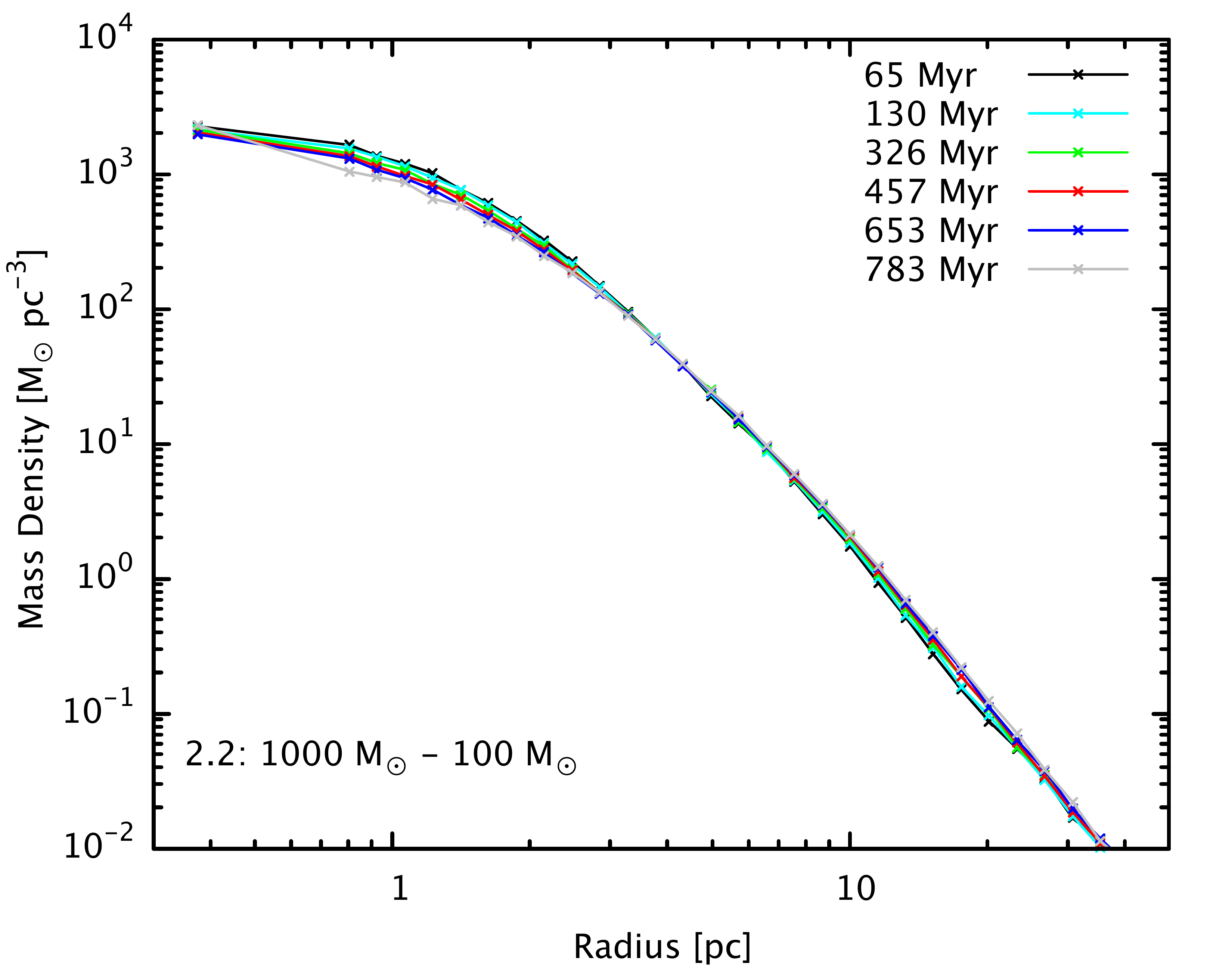}
	  \includegraphics[width=0.48\linewidth]{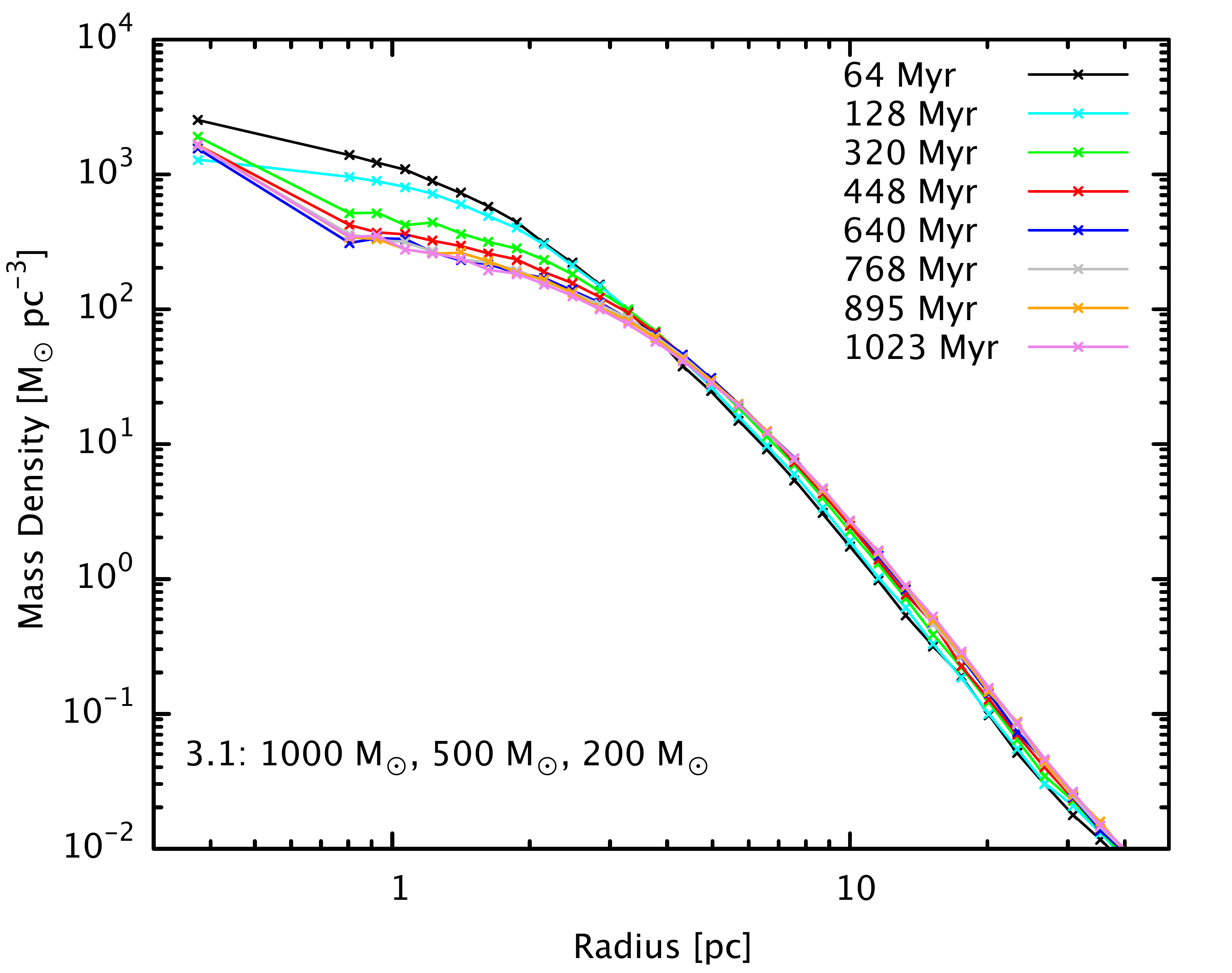}
	 \includegraphics[width=0.48\linewidth]{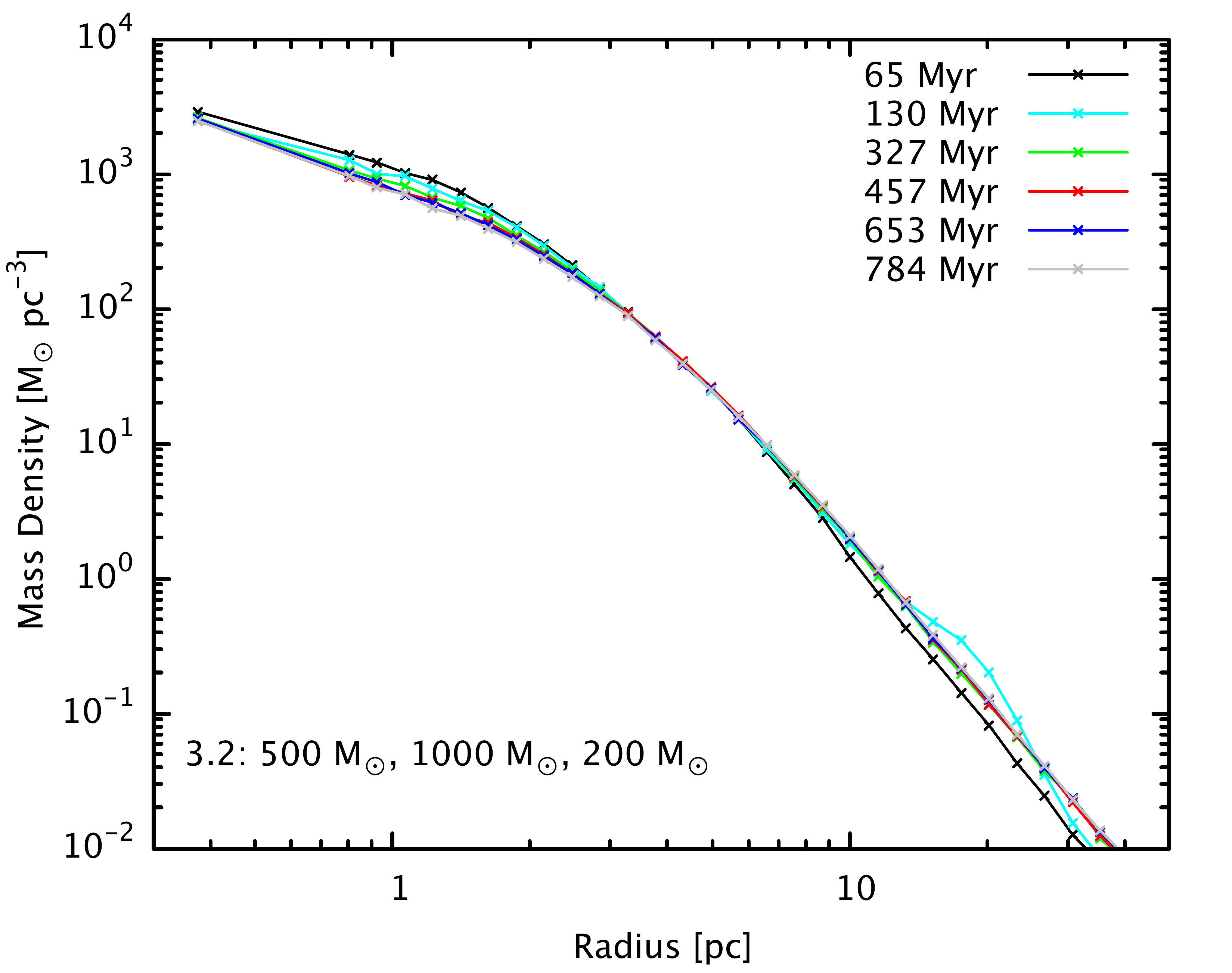}
	\caption{The top two panels show the evolution of the density profile of runs 2.1 (left) and 2.2 (right). The lower two panels show the evolution of the density profile for runs 3.1 and 3.2.}
    \label{fig:imbh-density-2-3}
\end{figure*}

In the case of the runs 2.i and 3.i, the formation of the IMBH binary leads to a slight decrease in the central density with time as surrounding stars are scattered away due to interactions with the IMBH binary as it gradually hardens (see Appendix \ref{ross-section}). The evolution and shape of the density profile is similar between runs 2.1, 3.2 and 3.3. The inner density profile (within 2 pc) for these runs is also steeper compared to runs 2.2 and 2.4 owing to the larger mass of the IMBH binary. Fig. \ref{fig:imbh-density-2-3} shows the evolution of the density profile for runs 2.1, 2.2, 3.1 and 3.2. The density profile for run 3.1 decreases steeply within the inner 2 pc after 400 Myr due to the scattering of the third IMBH (see Section \ref{subsec:strong3} and Fig. \ref{fig:3x-sep}).

Using the profiles shown in Fig. \ref{fig:imbh-density-0-1} and Fig. \ref{fig:imbh-density-2-3}. We fit the density profile to a power-law function given by $\rho \ (r) \propto r^{-\gamma}$. We find that that $\gamma$ ranges between 1.2 and 1.6 and tends to increase with the evolution time of the cluster. These values are consistent with observationally inferred density profiles of NSCs \citep{Neumayer2020,pechetti2020}.

\begin{figure*}
     \includegraphics[width=0.48\linewidth]{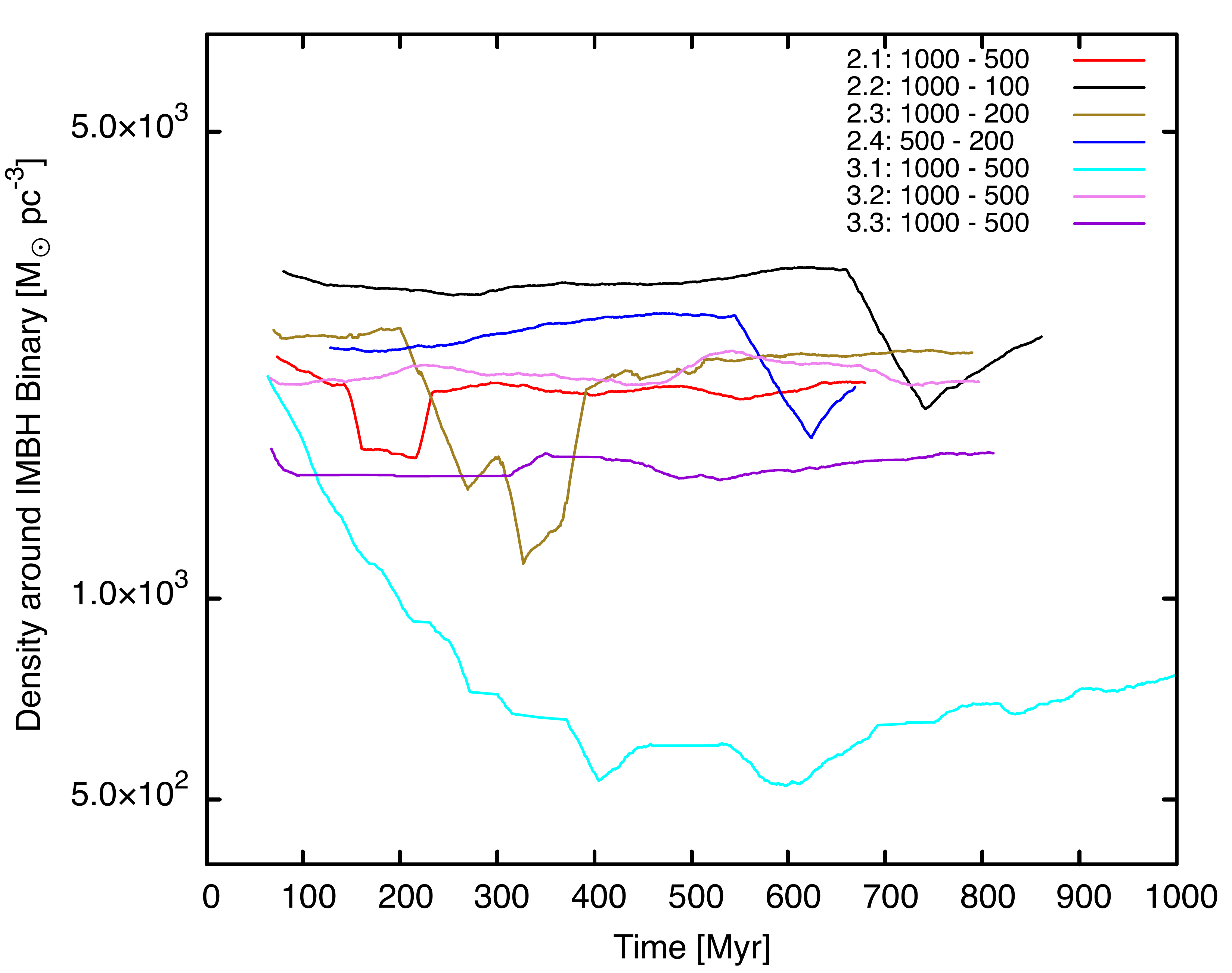}
	 \includegraphics[width=0.48\linewidth]{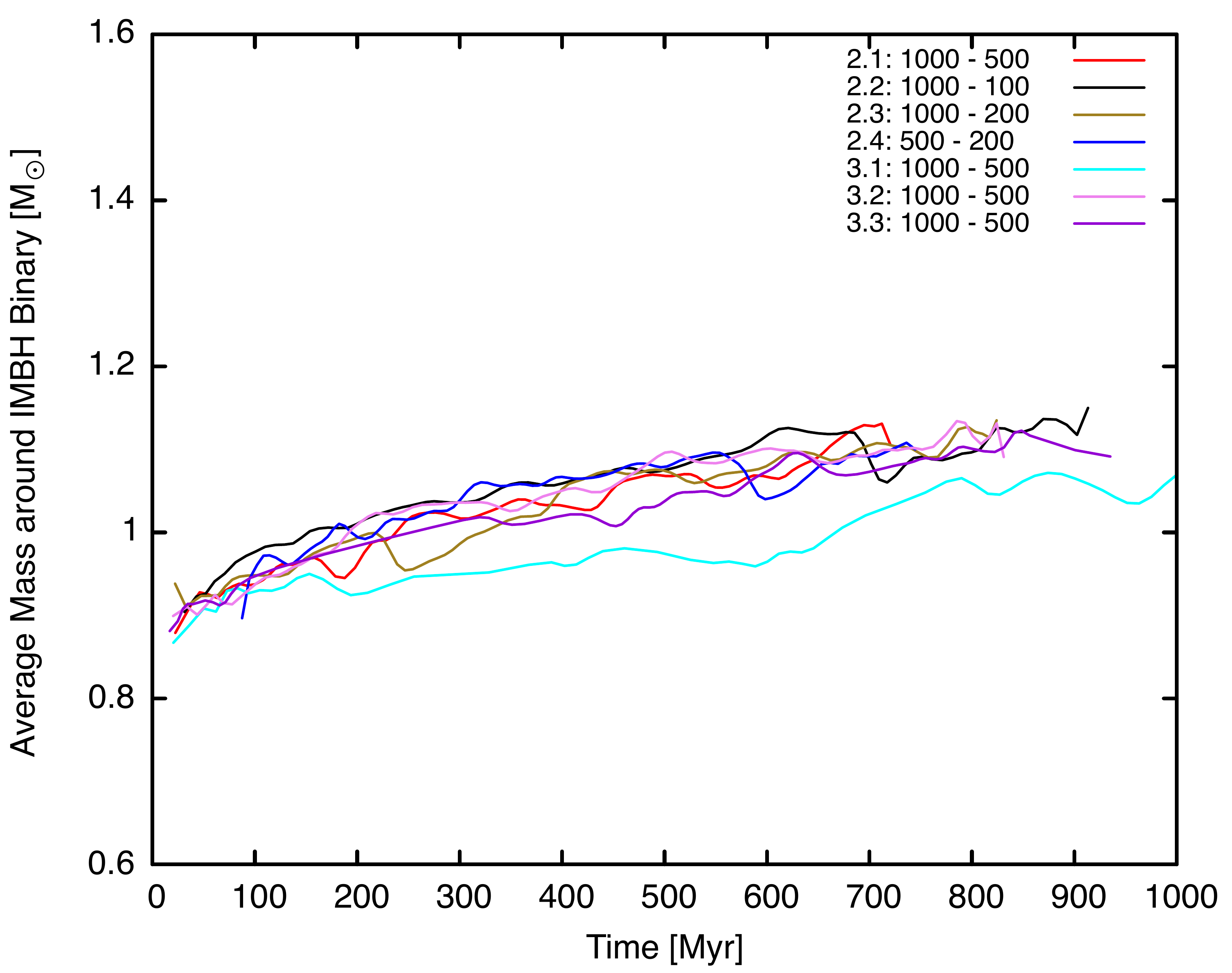}
	\caption{The left panel show the evolution of the density surrounding the IMBH binary in runs 2.i and 3.i. The right panel shows the average mass of the stars in the vicinity of the IMBH binary. The density and average stellar mass were calculated inside a box with length 0.5 pc centered on the centre of mass of the IMBH binary.}
    \label{fig:den-around-IMBH}
\end{figure*}

The left panel of Fig. \ref{fig:den-around-IMBH} shows the stellar density around the IMBH calculated within a radius of 0.25 pc. We find that the highest stellar density around the IMBH binary is for run 2.1 which also accounts for why its hardening is faster compared to the other runs (see Appendix \ref{ross-section}). We also find that the average mass surrounding the IMBH binary in all the 2.i and 3.i runs is gradually increasing with time. This can be seen in the right panel of Fig. \ref{fig:den-around-IMBH}. As the merged cluster evolves, more massive stars segregate closer to the centre \citep{panamarev2019} and the IMBH binary. Table \ref{tab:final-properties}, shows the approximate bound cluster mass (within $\sim$ 15 pc) and half-mass radius for the runs towards the end of the simulations. Stellar ejections also play a role in the cluster evolution and the evolution of the binary IMBH \citep{iwasawa2011,wang2014}. The bound cluster mass about 80 - 90 per cent of the initial cluster mass and is smaller for more evolved models and those with a more massive IMBH binary.

\begin{table}
\centering
  \caption{This table provides the estimated bound cluster mass and half-mass radius of the clusters close to end of the simulation.}
  \label{tab:final-properties}
\begin{tabular}{|l|c|c|c|}
\hline
\textbf{Run} & {\textbf{Time}} & {\textbf{Cluster Mass}} & \textbf{Half-Mass} \\ & [Myr] & [$\times 10^{4}$ $\rm M_{\odot}$] & \textbf{Radius [pc]} \\ \hline
0.1 & 375 & 7.7 & 2.30 \\
1.1 & 250 & 7.5 & 2.28 \\
1.2 & 357 & 7.5 & 2.30 \\
1.3 & 551 & 7.5 & 2.32 \\
2.1 & 716 & 7.2 & 2.76 \\
2.2 & 914 & 7.5 & 2.70 \\
2.3 & 825 & 7.5 & 2.61 \\
2.4 & 745 & 7.4 & 2.57 \\
3.1 & 1511 & 6.7 & 3.48 \\
3.2 & 832 & 7.1 & 2.68 \\
3.3 & 939 & 7.3 & 2.91
\end{tabular}
\end{table}

\section{The fates of IMBHs in merging stellar clusters}\label{imbh-removal}

In this section, we discuss the fates of IMBHs that are delivered to the centre of a galaxy. In the runs presented in Sections \ref{2i-runs} and \ref{sec4:3i-runs} we show that multiple IMBHs could potentially be delivered to the NSC. Whilst we expect that a single IMBH would always be retained in the NSC following delivery, two additional mechanisms can act if multiple IMBHs are present.

In both the 2.i and 3.i runs, we find that a binary IMBH forms and gradually hardens and becomes more eccentric. These binaries can efficiently merge through GW radiation on timescales of a few hundred Myr to a few Gyr. However, the merger will lead to a recoil kick from asymmetric emission of gravitational waves \citep{Fitchett1983,meritt2004}. This kick could remove the merger product from the NSC, leaving behind no seed black hole to grow into an SMBH. In Section \ref{sec:4-gw-recoil-kicks}, we discuss gravitational recoil kicks and the role they play in removing IMBH from the NSC.

Binary-single scatterings involving IMBHs could lead to the ejection of the single IMBH from the cluster, and possibly also the ejection of the binary. In Section \ref{sec4:binsin-interaction}, we discuss binary-single encounters involving three IMBHs, the possible outcomes, and their dependence on the properties of the interacting IMBHs. 

If an IMBH is retained in the NSC then it can seed the formation of an SMBH. In Section \ref{sec4:growth-smbh}, we discuss the main processes by which an IMBH could grow in an NSC in order to become an SMBH. In Section \ref{sec:caveats}, we discuss the caveats and limitations of the runs that we presented and how they may impact the results.

\subsection{GW recoil kicks and IMBH retention}\label{sec:4-gw-recoil-kicks}

In Sections \ref{2i-runs} and \ref{sec4:3i-runs}, we found that in all of our runs that contained two or more IMBHs, a binary IMBH forms within a few tens of Myr, which then gradually hardens and becomes more eccentric by scattering away stars (Appendix \ref{ross-section}). %
The merger time for a binary inspiralling due to GW radiation \citep{peters1964} is given by: 

\begin{equation}
\tau_{\rm gr} \simeq \ 10^{10} \ y r\left(\frac{a_{\rm bin}}{3.3 R_{\odot}}\right)^{4} \frac{1}{\left(m_{1}+m_{2}\right) m_{1} m_{2}} \cdot\left(1-e^{2}\right)^{7 / 2}
\label{eq:gr-merger}
\end{equation}

where $a_{\rm bin}$ is the semi-major axis of the binary, $e$ is the eccentricity, and the masses of the two black holes are $m_{1}$ and $m_2$ in solar mass.

As the eccentricity of the binary increases, the GW inspiral time given by Equation \ref{eq:gr-merger} decreases. We find that for runs 2.2 and 2.3, the binaries will merge within 500 Myr from the start of the simulation. Extrapolating, the evolution of the semi-major axis and eccentricity for the binary in runs 2.1, 3.2 and 3.3, we estimate the merger time to be around 1500 Myr. For run 2.4, we find a merger time of around 1000 Myr. The longest estimated merger time for the IMBH binary is 6000 Myr for run 3.1. Table \ref{tab:gw-merger-time} summarizes the estimated merger times for the different IMBH binaries in our runs. The extrapolation was done by fitting the evolution of the gravitational wave merger time calculated from the semi-major axis and eccentricity evolution from snapshots following the formation of the IMBH binary. As discussed in Section \ref{subsec:strong3}, following a strong interaction with a single IMBH, the binary eccentricity decreases and the subsequent recoil from the interaction places the binary in the part of the cluster where the stellar background density is low and thus the hardening rate for the binary decreases leading to a longer merger time. However, given that the timescale for hardening and eccentricity pumping depends on the density and velocity dispersion of stars around the binary IMBH, in a denser cluster the time needed for the binary to merge due to GW radiation may be a lot shorter \citep{rasskazov2019}. However, it may not be so straightforward to extrapolate our results to denser systems. This is discussed in more detail at the end of Section \ref{sec:caveats}.

\begin{table}
\centering
  \caption{This table provides the estimated merger time for the IMBH binaries in runs 2.i and 3.i.}
  \label{tab:gw-merger-time}
\begin{tabular}{|l|r|c|}
\hline
\textbf{Run} & {\textbf{IMBH Binary}} & \textbf{Estimated GW}  \\ &  \textbf{Mass [$\rm M_{\odot}$]} & \textbf{Inspiral Time [Myr]} \\ \hline
2.1 & $1000 \ \msun$ - $500 \ \msun$ & 1500 \\
2.2 & $1000 \ \msun$ - $100 \ \msun$ & 500\\
2.3 & $1000 \ \msun$ - $200 \ \msun$ & 500\\
2.4 & $500 \ \msun$ - $200 \ \msun$ & 1000\\
3.1 & $1000 \ \msun$ - $500 \ \msun$ & 6000\\
3.2 & $1000 \ \msun$ - $500 \ \msun$ & 1500\\
3.3 & $1000 \ \msun$ - $100 \ \msun$ & 1500
\end{tabular}
\end{table}

The GWs emitted by binary BHs as they inspiral are anisotropic. As the final inspiral occurs in less than one orbital period of the binary, the GWs carry away linear momentum. In order to compensate for this, the merged BH acquires a GW recoil kick  
\citep{Fitchett1983,Wiseman1992}.  The magnitude of this recoil kick depends strongly on the mass ratio of the merging BHs and their relative spin magnitudes and orientations
\citep{baker2007,baker2008,gonzalez2007,Campanelli2007,Lousto2008,Lousto2012}.For a perfectly symmetric binary system,
comprising equal mass non-spinning BHs, the recoil kick will be zero; however, large spin values with asymmetric orientations can lead to recoil kicks that can be of order of a few thousand $\kms$ \citep{holley2008,Lousto2012,blecha2016,fragione2018}.
Such kicks can eject the merged BH from its host NSC \citep{holley2008,Campanelli2007,morawski2018,gerosa2019,Dunn2020}. Mergers with more extreme mass ratios result in lower recoil kicks, and the merged remnant is more likely to be retained in dense environments with relatively large escape velocities \citep{morawski2018,antonini2019,rodriguez2019,fragione2020}.

\begin{figure}
	\includegraphics[width=\columnwidth]{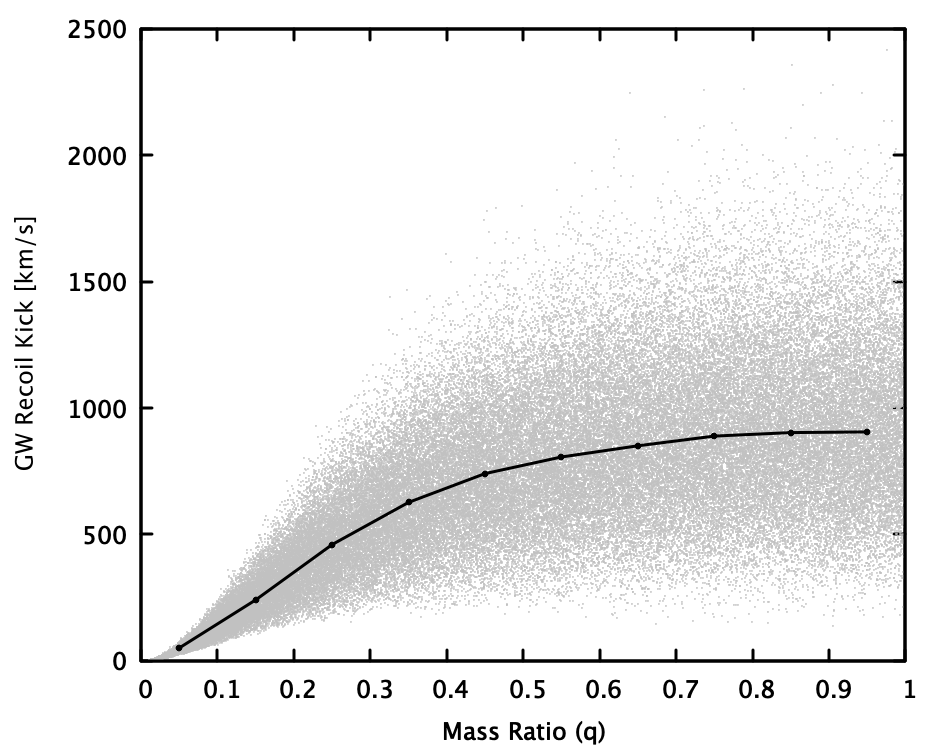}
	\includegraphics[width=\columnwidth]{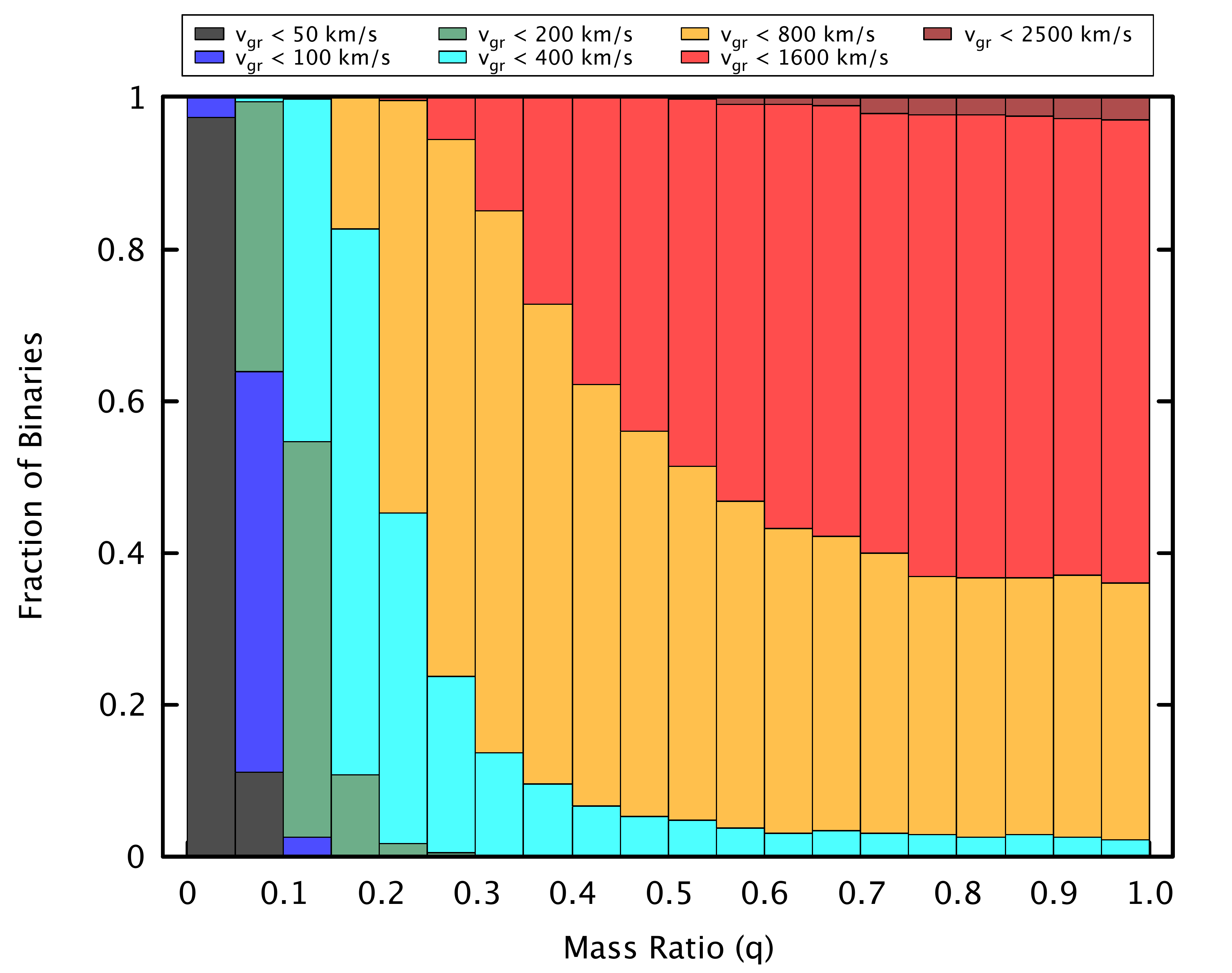}
	\caption{The top panel shows the gravitational wave recoil kick ($\rm v_{gr}$) as a function of the mass ratio (q) for $10^{5}$ merging BHs with a uniform mass ratio distribution. The black line shows the median value of recoil kicks, as a function of mass ratio. The lower panel histogram shows the fraction of binaries with recoil kick velocities lower than 50, 100, 200, 400, 800, and 2500 $\kms$ as a function of mass ratio.}
    \label{fig:recoil-kicks}
\end{figure}

In order to estimate the dependence of recoil kick velocities on the mass ratio of merging IMBH, we sampled a set of $10^{5}$ merging binary BHs with a uniform mass ratio distribution. A spin value was assigned to each BH by combining the `hot' and `cold' spin distributions provided in Fig. 7 of \citet{Lousto2012}. These spin distributions have a peak value of about 0.7 and take into account the accretion-driven growth of the BHs. Given that the binary IMBH in our clusters form dynamically, we assume an isotropic distribution of spin orientations. For each binary we then estimate the recoil kicks based on fits to results from numerical relativity simulations provided by \citet{baker2008,vanmeter2010} and average over five different spin orientations for each binary. The GW recoil velocity as a function of mass ratio is shown in Fig. \ref{fig:recoil-kicks}. We find that for the majority of the binaries with mass ratio $\rm q \lesssim 0.15$, the median recoil velocities would be of the order $\rm \lesssim 200 \, \kms$. In that case, the merged IMBH could be retained in NSCs where escape speeds are of the order of a few hundreds of $\kms$ \citep{antonini2019,gerosa2019}. For higher mass ratios of the merging BHs, GW recoil velocities can be much higher due to spin asymmetry leading to a significant contribution in the recoil kick velocity vector which is perpendicular to the orbital plane \citep{baker2008,vanmeter2010,Lousto2012}.

For our runs where the mass ratio of the binary IMBH is less than about $0.15$ (as in runs 2.2 and 2.3), the likelihood for the merged remnant to receive a low kick and be retained in a dense NSC will be higher. These are also the runs in which GW radiation merger times for the binaries are the shortest. For higher mass ratios, the likelihood 
for the merged BH to be ejected out of the NSC will increase. This process can result in the ejection of the IMBH from an NSC. Therefore, GW recoil kicks have an important consequence on whether or not an NSC can retain an IMBH that can go on to seed an SMBH \citep{meritt2004,volonteri2007}.

Similarly, GW recoil kicks are also important from the point of view of forming IMBHs in dense stellar clusters. If the initial seed BH forms from the collapse of a massive star created through runaway collisions and is of the order of a few hundred solar masses then the likelihood of retaining the seed following mergers with lower mass stellar BHs will be higher \citep{baker2008,vanmeter2010,morawski2018}. However, depending on initial spin distributions of BHs, slow growth of a low mass seed BH through mergers with other stellar-mass BHs of similar mass is likely to eject the seed from the cluster which will inhibit IMBH formation \citep{rodriguez2019,gerosa2019}.

\subsection{2+1 Scattering}\label{sec4:binsin-interaction}

Here we consider the effects of a binary-single encounter between three IMBHs and the consequences these interactions can have on the retention of IMBH in the merged cluster. Such encounters occur in our run 3.1 and could potentially result in the escape of one or multiple IMBHs. Here we estimate the recoil velocities that can come out of such interactions and under which circumstances could all three IMBHs be potentially removed from the NSC.

If we assume that we have a hard binary IMBH that has a binding energy which is much larger than the kinetic energy of the single IMBH that it encounters, then by conserving energy and momentum in the rest-mass frame of the encounter, we can estimate the resulting recoil speed of the binary to be

\begin{equation}
V_{\rm bin}=\sqrt{\alpha \frac{ m_{1} m_{2} m_{3}}{\left(m_{1}+m_{2}\right)\left(m_{1}+m_{2}+m_{3}\right)} \cdot \frac{1}{a}} \cdot 30 \ \mathrm{km} \ \mathrm{s^{-1}}
\end{equation}
where $m_{1}$ and $m_{2}$ are the masses of the binary components and $m_{3}$ is the mass of the single. The masses are in units of $\msun$ and $a$ is the semi-major axis of the binary in units of AU. $\alpha$ is the fractional change in the binding energy of the binary following the interaction, i.e, ${\Delta E_{b}}/{E_{b}}$. For a hard binary, $\alpha$ is proportional to the ratio between the mass of the single and the mass of the binary.

For the 3.i runs, the component masses of the binary IMBH are $ m_{1} = 1000 \, \msun$, and $ m_{2} = 500 \, \msun$, and the mass of the single IMBH is, $ m_{3} = 200 \, \msun$. Taking $\alpha=0.1$ \citep{hills1983}, the binary recoil velocity can be expressed in terms of the semi-major axis of the binary as 

\begin{equation}
V_{\rm bin}= 59.41 \ \mathrm{km} \ \mathrm{s^{-1}} \sqrt{\frac{1}{a}}
\end{equation}

In the context of runs 3.i, for the inner binary IMBH to receive a significant recoil kick ($>50 \, \kms$) in an interaction with the 200 $\msun$ IMBH, its semi-major axis should be at most 1.4 AU. For a higher $\alpha$ value of 0.3, this semi-major axis would be 4.2 AU. However, the 200 $\msun$ IMBH is more likely to be ejected out before the semi-major axis hardens to reach such a small separation and this is shown in all the 3.i runs. In runs 3.1 and 3.3, we find that the single IMBH is ejected from the cluster following a strong encounter with the binary IMBH at a time when the binary semi-major axis values are 50 AU and 200 AU.

If an additional IMBH is delivered to the NSC while the binary IMBH has hardened to about an AU, then a binary-single scattering could result in significant recoil velocity for the binary IMBH and this can potentially eject all three IMBHs from a NSC \citep{guletkin2006} with escape velocities of the order of several tens of $\kms$. However, according to Equation \ref{eq:gr-merger}, the GW radiation merger time for a binary IMBH with component masses of 1000 $\msun$ and 500 $\msun$, with a separation of 1 AU and an eccentricity of 0.9, is around 1 Myr. 
This merger timescale for the binary IMBH is much shorter than the expected injection frequency of an additional IMBH into the NSC which may occur on timescales of about a 100 Myr to up to 1 Gyr.
Therefore, the likelihood of a single IMBH to encounter the binary IMBH while it is hard enough to get a significant recoil velocity is small. The most likely outcome of an interaction between the binary IMBH and a single IMBH is either ejection of the single IMBH or its exchange with one of the binary components.

Interactions between a binary IMBH and another IMBH can influence the eccentricity evolution of the binary described in Appendix \ref{ross-section}.  In run 3.1, we found that the merger time for the binary IMBH becomes longer following an interaction with the third IMBH which leads to ejection of the latter. This interaction decreases the eccentricity of the binary which increases its merger time. In runs 3.i, all three IMBH were there from the very beginning. However, if the third IMBH is delivered with a merging cluster after a few hundred Myr of the formation of the binary IMBH then it is likely that the binary IMBH has already hardened and become sufficiently eccentric that it would merge before it encounters the third IMBH (see Section \ref{sec4:binsin-interaction}). Other works have investigated triple interactions between IMBH/SMBH in nuclei of galaxies \citep{hoffman2007,sedda2019b} and found that such encounters can lead to erratic changes in the orbital eccentricity of the binary which could drive their prompt merger. Additionally, it has also been found that such interactions could lead to the ejection of BHs from the nuclei to the halo of the galaxy.

\subsection{Growing seeds to super-massive black holes}\label{sec4:growth-smbh}

We argue that if an NSC forms through mergers of stellar clusters then it is possible that some of these merging clusters may bring along IMBHs. These IMBHs can potentially grow in the galactic nuclei in order to become SMBHs. However, if the merging clusters do not bring along an IMBH then the nuclei will not contain a seed BH that can grow on to become an SMBH.

In run 0.1, we showed that the merger of three stellar clusters without any IMBH can lead to the formation of a merged cluster with a dense core. Relatively low mass galaxies that have NSC masses of the order of few $10^{6} \, \msun$ (like M33) may have formed their NSC through the merger of just a few stellar clusters. It is probable that none of these merging clusters contained an IMBH and therefore no IMBH is delivered to the galactic nuclei in these galaxies.
If one IMBH is delivered to a galactic nucleus (runs 1.i) then the IMBH can stay in the nucleus where it can grow by either accreting gas or tidally disrupting stars \citep{Devecchi2009,Davies2011,stone2017,fragione2018b,natarajan2020}. The stellar clusters that merge to form an NSC may also contain gas that could be be delivered to the galactic nuclei \citep{Davies2011,guillard2016}. Additionally, infall of gas into an NSC can also occur during galaxy mergers \citep{mayer2010}. Gas in the NSC can be efficiently accreted by the IMBH \citep{leigh2013,inyoshi2016,sakurai2017,natarajan2020} and can result in the growth of the NSC by triggering in-situ star formation \citep{Neumayer2020}. Such inflow of gas will also effectively contract the NSC \citep{Davies2011}, leading to higher central densities which would increase the rate of tidal disruption events.

\citet{Miller2012} argued that in dense NSCs with velocity dispersions larger than $40 \, \kms$, binaries will be unable to support against core collapse which will lead to the runaway growth of BHs. \citet{stone2017} showed that in such NSCs, stellar BHs can become more massive by growing through tidal capture and disruption of stars. These tidal disruptions events can effectively grow the mass of a BH mass from $10^{2-3} \, \msun$ to up to $10^{5} \, \msun$ \citep{stone2017,alexander2017,sakurai2019,seguel2020}. IMBH can also grow by tidally disrupting white dwarfs \citep{rosswog2009,macleod2016}. In addition to gas accretion, these dynamical processes can lead to the growth of the seed SMBH to larger masses. Furthermore, if the NSC grows through additional mergers of stellar clusters or gas accretion this could then replenish the environment around the seed SMBH.

In our simulated runs with two or more IMBH, we find that a binary IMBH forms. The binary IMBH scatters away stars and becomes more eccentric with time (Appendix \ref{ross-section}). Due to this increased eccentricity, the binary can efficiently merge due to GW radiation.  As discussed in Section \ref{sec:4-gw-recoil-kicks}, after such a merger, the merged IMBH can receive a significant recoil kick which can be up to a few thousand $\kms$ and can eject the IMBH from the NSC. This can remove the seed SMBH from the merged galaxy. However, if the recoil kick velocity is less than the escape speed of the NSC then the merged IMBH can be retained. The recoil velocity is lower for merging BHs with more extreme mass ratios. Therefore, if the initial seed IMBH in an NSC can grow its mass through gas accretion and/or tidal disruption of stars then the likelihood of retaining the seed following subsequent mergers with lower mass IMBHs will be higher \citep{blecha2011}. These mergers can also contribute to the growth of the seed SMBH in the NSC.

\subsection{Caveats and limitations}\label{sec:caveats}

In this subsection, we discuss the main limitations of our simulated runs and discuss how those limitations may influence the results presented in this paper. 

For our merging stellar clusters, we use a limited mass function with stars not more massive than 2 $\msun$. The main sequence turn-off age for 2 $\msun$ stars is about 800 Myr for a metallicity of $Z=0.001$. Up to few hundred Myr of cluster evolution may be needed to form an IMBH of around a 1000 $\msun$ in the individual cluster models that were merged together and by this time stars with a mass larger than 3 $\msun$ would have already evolved. Therefore, the average mass of stars would still be significantly smaller than the mass of the IMBHs in the cluster. It can be expected that many of the stellar-mass BHs that form in these clusters are ejected due to natal kicks or have merged with the IMBH \citep{Lutz2013,leigh2014,Giersz15,hong2020}. However, it is possible that there may still be few tens to hundreds of stellar mass BHs retained in the cluster at the time when the clusters fall in and merge. The inclusion of stellar-mass BHs in the initially merging clusters could potentially influence the eccentricity evolution of the binary IMBH discussed in Appendix \ref{ross-section} through strong dynamical interactions with the binary IMBH. Additionally, the presence of an IMBH in a stellar cluster can also result in mergers of stellar BHs binaries through Lidov-Kozai oscillations \citep{haster2016,frag2019,martinez2020,samsing2018}. We will investigate the influence of including stellar-mass BHs within our merging stellar cluster models in a future study.

For simplicity, we did not consider any primordial binaries in our merging cluster set-up. Stars in primordial binaries may not have a significant influence on the segregation and formation of a binary IMBH. However, stellar-mass BHs in binaries could strongly interact with IMBHs as they segregate to the cluster centre. The \textsc{MOCCA} models described in Section \ref{subsec:imbh-formation} that do form an IMBH contain both primordial binaries and BHs. These BHs either merge with the IMBH or are ejected out of the cluster through dynamical interactions \citep{Giersz15}. 

Additionally, our merged stellar clusters are not of the same size or density as NSCs observed in galaxies like M33 or the Milky Way. The estimated upper limit for mass of the NSC in M33 is $2 \times 10^{6} \, \msun$ \citep{kormendy1993,graham2009} and \citet{gordon1999} found the mass of the M33 NSC to be about $7 \times 10^{5} \, \msun$. To keep a manageable number of stars in our simulated runs, we had scaled down the initial mass of the merging clusters and their initial densities were of the order of $\rm 10^3 \, \msun \, pc^{-3}$. It can be expected that the clusters merging with the galactic centre will be more massive and dense \citep{hartmann2011,Antonini2012}. Having more dense and massive clusters with larger escape velocities will increase the likelihood of those clusters forming and retaining an IMBH \citep{antonini2019}. Higher stellar density and velocity dispersion may also increase the hardening rate \citep{rasskazov2019}, hence decreasing the time needed for the binary IMBH to merge due to GW emission. Additionally, it would also lead to faster growth of the seed SMBH in the NSC \citep{stone2017} as discussed in Section \ref{sec4:growth-smbh}. However, there are several effects that may affect the binary evolution in denser systems. Firstly, dynamical friction would be less efficient and the IMBH in-spiral phase may last longer. Secondly, the hard binary separation would be smaller and therefore the binary separation would need to shrink sufficiently before stellar hardening becomes dominant. Thirdly, the influence radius of the IMBH binary would be much smaller and this would influence the orbit of stars surrounding the IMBH binary. This can imply  less efficient refilling of the binary loss-cone and may lead to less efficient hardening of the IMBH binary. Lastly, the models presented here do not account for the tidal field exerted from the inner galaxy ($\lesssim 100$ pc) onto the infalling clusters, which can significantly strip out the outer layer of the smaller clusters and potentially reduce the capability of the cluster nucleus to reach the innermost regions of the galactic centre.

\section{Conclusions}\label{sec:conclusions}

In this paper, we have considered the idea that an NSC forms first and that the SMBH grows later. In this scenario, the formation of an NSC occurs through mergers of stellar clusters in a galactic nuclei. Some of these clusters may have formed IMBHs in them through dynamical processes described in Section \ref{subsec:imbh-formation}. As the clusters merge, those IMBH(s) are delivered to the galactic nuclei where they can potentially grow and become seed SMBHs \citep{ebisuzaki2001}.

In order to investigate the viability of this scenario, we simulated the final stages of an idealized merger of three stellar clusters to form a merged cluster using \textit{N}-body simulations. We consider a range of cases where various combinations of the three merging clusters can contain IMBHs. In our simulated runs, we have a hierarchy of masses in the merging stellar clusters: the mass of the IMBH (100-1000 $\msun$) is smaller than the mass of the three merging stellar
clusters (up to $8 \times 10^{4}$ $\msun$) and the masses of stars 
(0.5 to 2.0 $\msun$) in our cluster are smaller than the IMBH mass ($\rm M_{\star} \ll M_{imbh} \ll M_{\rm cluster}$). Additionally, there is no pre-existing SMBH in our simulations.

Since $\rm M_{imbh} \ll M_{\rm cluster}$, the properties of the merged cluster are not significantly influenced by the presence of an IMBH. For runs that do contain an IMBH, the timescale for segregation of the IMBHs to the merged cluster centre is of the order of a few tens of Myr. Additionally, given that $\rm M_{imbh} \gg \rm M_{\star}$,  this leads to efficient segregation of the IMBH(s) to the centre of the merged cluster due to dynamical friction.

For our runs with two or more IMBHs, we find that the IMBHs sink to the centre of the merged cluster (within 
30 - 80 Myr) resulting in the formation of a binary IMBH. This binary hardens by scattering stars and becomes significantly eccentric in the process. Due to this increase in eccentricity, the time needed for the binary to merge due to GW emission decreases significantly. We found that the time needed for the merger of the IMBH binary ranges from several hundred Myr to a few Gyr. As discussed in Section~\ref{sec:4-gw-recoil-kicks}, the retention of the merged IMBH will depend on the magnitude of the GW recoil kick. These kicks depend on the mass ratio of the merging BHs, and their spin magnitudes and orientations. We argue that the merged BH can be retained in typical NSCs provided that the mass ratio of the merging IMBH is low ($\rm q \lesssim 0.15$), otherwise gravitational recoil kicks can be larger than a few hundred $\kms$ and can remove the merged BH from the NSC. Additionally, if the binary IMBH dynamically interacts with a third IMBH within the NSC then this can lead to the ejection of the third IMBH or, in an unlikely case, the ejection of all three IMBHs. Given these different outcomes, NSCs constructed through merging clusters will either contain no seed BHs or they will retain a single seed IMBH that can further grow to become an SMBH. 

This scenario would naturally explain the absence of SMBHs in galaxies like M33, where there was either no seed SMBH in the NSC to begin with or the product of the merger of two IMBHs was ejected due to GW recoil kicks. In massive galaxies with more massive NSCs, a higher number of clusters could have merged with the NSC which would increase the likelihood of delivering an IMBH which could then grow on to become an SMBH. In a companion paper, we model the build up of NSCs for a population of galaxies by merging stellar clusters. In these NSCs, we consider the retention and subsequent growth of seed SMBHs through GW mergers of IMBHs and subsequent gas accretion. Moreover, we check the consequences of these results on the observed demographics of NSCs and the occupation fraction of SMBHs in galaxies.

\section*{Acknowledgements}

We would like to thank the reviewer for a comprehensive report that helped in improving the quality and readability of the manuscript. We would like to thank Long Wang for his assistance in using \textsc{NBODY6++GPU} and to the developers of the code for making it publicly available. AA was supported by the Carl Tryggers Foundation for Scientific Research through the grant CTS 17:113. RC and AA acknowledge support from the Swedish Research Council through the grant 2017-04217. We would like to thank the Royal Physiographic Society of Lund and the Walter Gyllenberg Foundation for the research grant: `Evolution of Binaries containing Massive Stars'. Simulations in this project were carried out at LUNARC, which is the centre for scientific and technical computing at Lund University through the projects LU 2018/2-28, 2019/2-27 and LU 2020/2-14. Simulations were also carried out using the `chuck' computer cluster hosted at the Nicolaus Copernicus Astronomical Centre (CAMK) in Warsaw, Poland.

\section*{Data Availability Statement}

The simulations in this project were carried out using \textsc{nbody6++gpu}. The data and output from these simulations will be shared on request to the corresponding author.

\bibliographystyle{mnras}
\bibliography{biblio} %

\appendix

\section{Dynamically-driven evolution of binary IMBHs}\label{ross-section}

A binary IMBH in one of our simulations forms at moderate to high eccentricities, immersed in a sea of lower-mass stars. As it interacts with those stars it loses energy and the binary's orbit hardens, but it also either becomes more eccentric or maintains a high eccentricity (see Fig.~\ref{fig:2x-binary}).  Other authors have observed this phenomenon of eccentricity-pumping; e.g. \citet{amaro2006,SeddaAMB2019} in IMBH binaries, \citet{Rastello2019} in stellar mass BH binaries and \citet{sesana2010,iwasawa2011,khan2012,Ogiya2020,Bonetti2020} in SMBH binaries. In each case the distinguishing feature is that the bodies that make up the binary are more massive than the bodies with which the binary is interacting.  We investigated to see whether this is a general effect.

Our binaries are, at least initially, too wide for general relativistic effects to play a significant role in their orbital evolution.  Hence changes of semi-major axis $a$ and eccentricity $e$ are driven by interactions with stars.  Such interactions can be with the smooth potential arising from the combined effects of many stars; alternatively, strong scatterings with individual stars can also change the binary's orbit.  It is the second effect that we consider here.

The binary IMBH that we form is, within a short time frame, dynamically hard (strongly bound). This means that it cannot be disrupted by an encounter with an individual star, and that the net effect of encounters is to harden the binary further.  It is convenient to analyse the effect on the binary orbit by considering the interaction between one of the two IMBHs and a star.  These encounters systematically act in the opposite direction to the relative motion of the star and the IMBH.  Since the binary orbital velocity $v_{\rm orb} \gtrsim \sigma$, where $\sigma$ is the stellar velocity dispersion, this, on average, slows the IMBH along its original direction of motion.  It also imparts a small additional momentum in a direction perpendicular to the IMBH's original direction of motion, the plane of which depends on the initial velocity of the incoming star.  

We consider for now solely the braking component of the encounter and analyse its effect on the orbit at pericentre and apocentre.  At pericentre, slowing the motion of one of the stars does not affect the pericentre separation, $r_{\rm peri}=a(1-e)$, since the resulting orbit is closed, but decreases $a$.  Hence braking encounters at pericentre circularise an orbit, as seen in, for example, the effects of tides on binary star orbits.  Conversely, braking encounters at apocentre do not change the apocentre separation, $r_{\rm apo} = a(1+e)$, but decrease the semi-major axis.  This drives the orbit to higher eccentricities.  This means that for purely braking encounters, the ratio of the rate of encounters around pericentre to that around apocentre will determine whether the orbit is driven to lower or higher eccentricities.  For a binary embedded at the centre of a uniform-density sea of low-mass particles, as in Chandrasekhar's derivation of dynamical friction, the binary spends longer time at apocentre than pericentre, and hence more encounters happen at apocentre.  Hence as long as $a$ is smaller than the scale on which the stellar density changes we would expect encounters with low-mass stars to drive the binary to higher eccentricities.

However, as Fig.~\ref{fig:2x-sep} shows, this is not an appropriate description of the physical scenario in which our binaries find themselves.  After a short space of time they have scattered out all the stars within their orbits.  Further orbital evolution is driven by stochastic encounters with stars that are scattered in from outside.  Hence we construct a toy model of these scatterings to investigate whether it behaves like our $N$-body simulations.  We draw stars from the same mass function as in our simulations, a power-law with index $\gamma=-2.3$ between $0.5\,\msun$ and $2\,\msun$.  Following an inspection of the kinematic properties of the stars within 0.1\,pc of the central binary in our $N$-body models, we take the stars to have a mass density $\rho = 10^4\,\msun\,{\rm pc}^{-3}$. Their velocities follow a Maxwellian distribution with $\sigma=15\,\kms$.  We choose impact parameters $b$ distributed uniformly in $b^2$ up to $b_{\rm max} \equiv \tilde b_{\rm max}a$ with $\tilde b_{\rm max}=100$.  The time between encounters is exponentially distributed with a mean interval of
\begin{equation}
\bar{\Delta t} = \frac{1}{n \sigma \uppi \tilde b_{\rm max}^2 a^2}.
\end{equation}

We consider encounters between the star and one of the two BHs, chosen at random. For each encounter we choose a uniformly distributed mean anomaly for the central binary and  solve the Kepler problem to obtain the position of our chosen IMBH in the binary centre of mass frame.  The encounter between the star of mass $m$ and the IMBH of mass $M$ acts to change the velocity of the IMBH parallel and perpendicular to the relative velocity of the IMBH and star following \citep[][eq.~3.54]{BinneyTremaineBook}
\begin{equation}
\left|\Delta \mathbf{v}_{M\perp}\right| = \frac{2mV_0}{M+m}\frac{b/b_{90}}{1+\left(b/b_{90}\right)^2}
\end{equation}
and
\begin{equation}
\left|\Delta \mathbf{v}_{M\parallel}\right| = \frac{2mV_0}{M+m}\frac{1}{1+\left(b/b_{90}\right)^2}
\end{equation}
where the relative velocity is $V_0$ and 
\begin{equation}
b_{90} = \frac{G(M+m)}{V_0^2}.
\end{equation}
position of IMBH in binary COM frame.

Since the stars should approach the binary with isotropically distributed velocity vectors we 
pick a perpendicular direction uniformly distributed on a circle.  We then apply the effects of the encounter to the IMBH's velocity vector, recompute the orbital elements of the binary, and repeat.

\begin{figure}
    \begin{center}
        \includegraphics[width=\columnwidth]{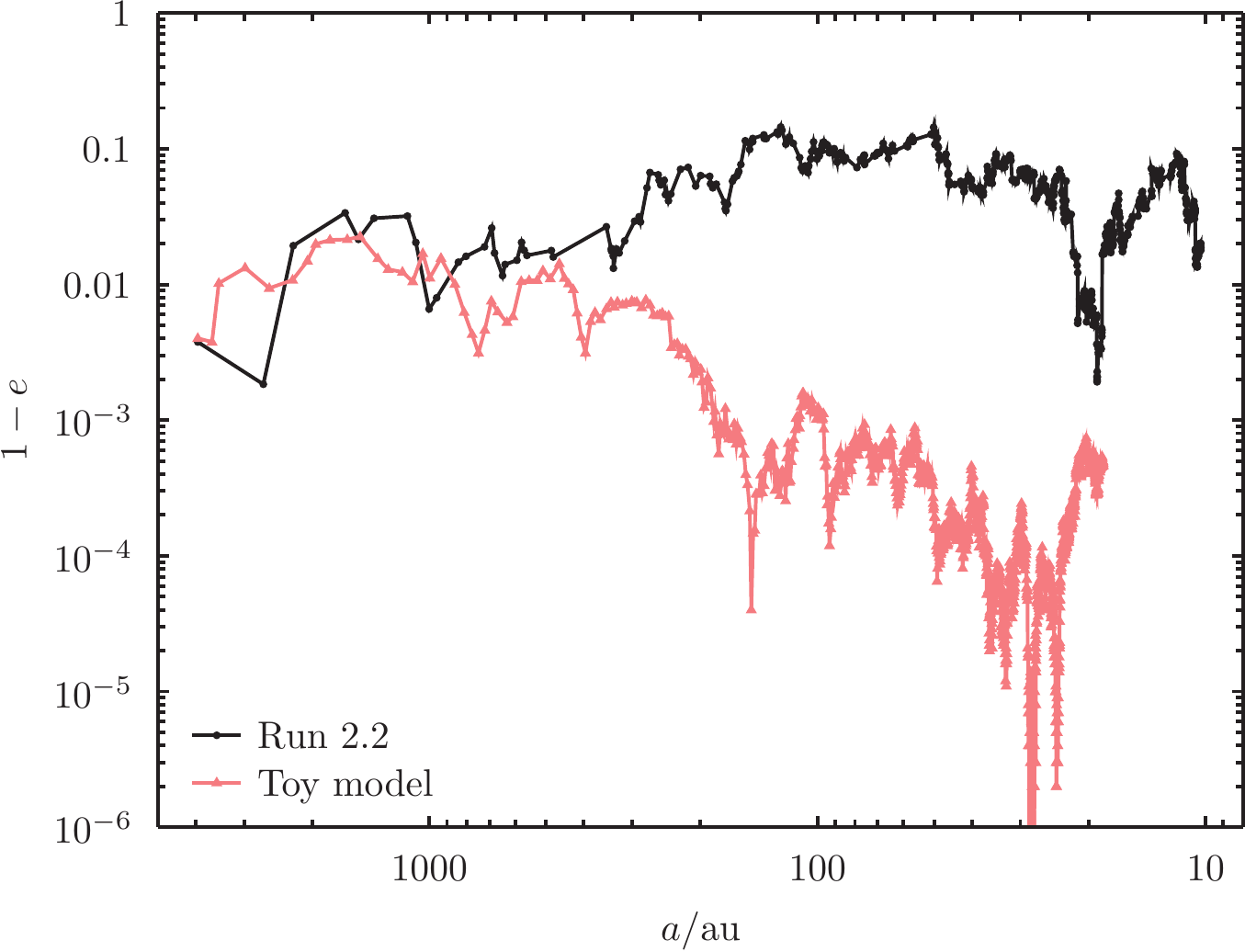}
        \caption{
            Comparison of toy model with $N$-body model.  The plot shows the semi-major axis $a$ vs $1-e$ up to a time of 763\,Myr, starting when the binary formed.  The black line shows results from the
            $N$-body model for binary 2.2 and the pink line shows the results from the toy model.
        }
        \label{fig:kickRepeatedly}
    \end{center}
\end{figure}

The resulting evolution of the orbital elements for our model of run 2.2 are shown in Fig.~\ref{fig:kickRepeatedly}.  The model qualitatively captures some of the key features of the evolution.  In roughly the correct time span the binary is driven together, with the orbital semi-major axis shrinking by a factor of about 300.  At the same time the binary is maintained at very high eccentricities, but the eccentricity fluctuates considerably as the IMBH, almost stationary at apocentre, is buffeted by the incoming stars.  There are two deficiencies to the toy model.  The eccentricity is still driven to higher values than in the $N$-body run, and the orbit does not shrink enough.  Both of these differences can be attributed to only considering the effect of the encounters on one of the IMBHs at a time.  In reality, weak encounters at large impact parameters torque both stars, systematically removing more energy from the binary and reducing the impact on the eccentricity, since the two IMBHs will be scattered in roughly the same direction.  However, the degree of qualitative agreement that we do see in Fig.~\ref{fig:kickRepeatedly} confirms that the eccentricity pumping is a physical effect owing to the processes that we describe and not a manifestation of numerical problems in the $N$-body code. Additionally, in the \textit{N}-body model the binary is free to move, compared to the toy model. This implies that the binary recoils in scattering events, thus part of the energy transfer is absorbed by the binary motion rather than being fully dissipated into the scattered star. As a result the binary hardening is less effective in the simulated run as compared to the toy model.

\bsp	%
\label{lastpage}
\end{document}